\documentclass[a4paper,fleqn]{cas-sc}

\usepackage[authoryear,longnamesfirst]{natbib}

\def\tsc#1{\csdef{#1}{\textsc{\lowercase{#1}}\xspace}}
\tsc{WGM}
\tsc{QE}

\begin{document}
\let\WriteBookmarks\relax
\def\floatpagepagefraction{1}
\def\textpagefraction{.001}

\shorttitle{}    

\shortauthors{}  


\title [mode = title]{The prospect of a fair outcome triggers prosocial behaviour in the Centipede Game}




\author[1]{Marco Saponara}[orcid=0009-0008-8092-6220]

\cormark[1]


\ead{marco.saponara@ulb.be}


\credit{Conceptualization, Methodology, Formal analysis, Software, Visualization, Writing - original draft}

\author[2]{Ant\'onio M. Fernandes}[orcid=0000-0002-4717-7984]
\ead{antonio.m.fernandes@tecnico.ulisboa.pt}
\credit{Conceptualization, Methodology, Writing – review \& editing}

\author[1,3]{Elias {Fern\'andez Domingos}}[orcid=0009-0009-1976-6202]
\ead{elias.fernandez.domingos@ulb.be}
\credit{Conceptualization, Methodology, Writing – review \& editing}

\author[2]{Ana Paiva}[orcid=0000-0003-3431-8060]
\ead{ana.paiva@inesc-id.pt}
\credit{Supervision, Writing – review \& editing}

\author[1,3,4]{Tom Lenaerts}[orcid=0000-0003-3645-1455]
\ead{tom.lenaerts@ulb.be}
\credit{Supervision, Funding acquisition, Conceptualization, Methodology, Writing – review \& editing}


\affiliation[1]{organization={Machine Learning Group, Université Libre de Bruxelles},
            city={Brussels},
            postcode={1050}, 
            country={Belgium}}

\affiliation[2]{organization={INESC-ID and Instituto Superior Técnico, Universidade de Lisboa},
            city={Lisbon},
            postcode={1000-029}, 
            country={Portugal}}

\affiliation[3]{organization={Artificial Intelligence Lab, Vrije Universiteit Brussel},
            city={Brussels},
            postcode={1050}, 
            country={Belgium}}

\affiliation[4]{organization={Center for Human-Compatible AI, UC Berkeley},
            city={Berkeley},
            state={CA},
            country={USA}}

\cortext[1]{Corresponding author}

\fntext[1]{}
\tnotemark[1,2]

\tnotetext[1]{The authors gratefully acknowledge the research support of the F.R.S-FNRS (project grant 40007793). T.L. further acknowledges the support of the Service Public de Wallonie Recherche (grant 2010235-ARIAC) by DigitalWallonia4.ai and the Flemish Government through the AI Research Program. E.F.D. is supported by an F.W.O. Senior postdoctoral grant (12A7825N).
A.F and A.P. are  supported by the INESC-ID (UIDB/50021/2020) and the Centre for Responsible AI (CRAI) project (grant no. C645008882-00000055/510852254 and C628696807-00454142, IAPMEI/PRR).}

\tnotetext[2]{The authors also thank Eladio Montero for his useful comments and suggestions, which allowed us to improve this article.
}


\begin{abstract}
Human decision-making reflects not only material incentives, but also concern for others’ gains and the ability to understand others.
However, the extent to which sensitivity to fairness interacts with aspects of social cognition and affects sequential social interactions requires further understanding.
Here, we address this question through a behavioural experiment on sequential exchanges of resources, represented by the Incremental Centipede Game (ICG).
The design comprises a `zero-end' treatment ($n=164$), where the ICG has a deadline yielding no payoff, and a `fair-end' treatment ($n=162$), where the last interaction leads to a fair outcome. 
The ICG, which is repeated twice, is complemented by two individual tasks: the Social Value Orientation task (SVO), to elicit social preferences, and the Cognitive Reflection Test (CRT), to assess participants’ deliberation.
Our results show that the fair outcome, though uncertain, prompts participants to move beyond self-interest. 
Indeed, an evolutionary model shows that the fair split shifts play towards a mutually beneficial strategy when players are sufficiently prosocial and use probabilistic strategies. 
A second game repetition reinforces this pattern, with adaptation towards cooperation. Although CRT scores do not predict behaviour in the ICG, participants who deviate from game-theoretic rationality exhibit stronger prosociality in the SVO. 
Notably, in the fair-end treatment, some non-rational players are individualists, suggesting that the fair split may function as a nudge.
In conclusion, our findings suggest that certain personal traits, such as social preferences, allow individuals to overcome self-interest to better align with their peers and achieve better collective outcomes in mixed-motive games.
\end{abstract}


\begin{highlights}
\item A fair but unsure outcome prompts humans to go past self-interest in a Centipede Game 
\item An evolutionary model with noisy prosocial agents explains the behavioural data
\item Fairness-seeking players increase in the second iteration of the Centipede Game
\item Players who deviate from game-theoretic rationality exhibit stronger prosociality
\end{highlights}


\maketitle

\section{Introduction}\label{sec:introduction}

Empirical evidence on strategic interactions reveal a pronounced variation in human social motives. 
For some, behaviour is mainly guided by personal material gains~\citep{doi:10.1126science.1110600}. 
For others, considerations of fairness and collective welfare shape their decisions, even when doing so entails a personal loss~\citep{grund2013natural, 10.1257.jel.20241391}.
Several theories have been proposed to explain the emergence of human prosociality, including kin selection, direct and indirect reciprocity, and norms~\citep{kurzban2015evolution, fletcher2009simple}.
More recently, research has argued that prosocial and altruistic behaviours may have originated from the ability to understand the beliefs and intentions of others, and to feel their emotional states~\citep{artinger2014others, singer2014understanding, 10.1098.rstb.2015.0097}.
These abilities, commonly referred to as \emph{theory of mind} and \emph{empathy} respectively, have been crucial to the formation of larger societies, as they facilitate the alignment among peers to achieve a collective goal~\citep{tomasello2005understanding, doi:10.1073.pnas.0601428103}.
In fact, to reduce the chances that someone else exploits their prosocial tendency, cooperative individuals need more accurate beliefs than individualists, who are likely to defect regardless of their beliefs about the others' intention~\citep{kelley1970social, mckay_evolution_2009}.

Although these cognitive mechanisms have been extensively investigated in isolation, the interplay between sensitivity to fairness, social cognition, and strategic decision-making requires further understanding~\citep{tusche2021neurocomputational}. 
To bridge this gap, we address three questions: 
(i) how and to what extent does the prospect of a fair future outcome affect social interactions?
(ii) to what extent can behavioural differences be ascribed to different degrees of social orientation?
(iii) what do observed choice patterns imply about the cognitive processes underlying strategic decision-making?
Here, we investigate them through a preregistered behavioural experiment on strategic decision-making in the context of sequential interactions. 
This setting is particularly interesting for our objective, as it encourages participants to think about how their own actions influence the subsequent choice of the co-player.
In addition, participants could engage in further recursive mechanisms (i.e., holding mental representations of their partners’ mental representations) that ultimately lead to interdependent cognitive processes~\citep{rusch_theory_2020}.

\begin{figure}[!ht]
\centering
\resizebox{\textwidth}{!}{%
\includegraphics{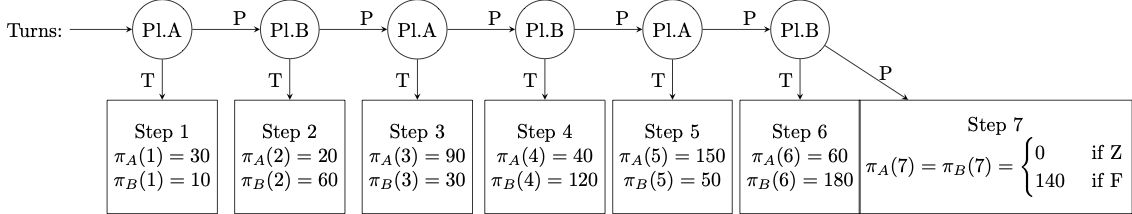}
}
\caption{Extensive form of the two Incremental Centipede Games used in our experiment. 
Player A (B) plays at odd (even) steps.
The shared resource starts at 40 and increases linearly for six turns.
Playing $T=$\textit{Take} at a given step means ending the game and receiving $75\%$ of the resource, while the co-player receives the remaining $25\%$.
Playing $P=$\textit{Pass} means letting the other player decide what to do in the next step, unless the last decision step is reached, where Player B has to decide between two different splits. 
The treatment determines the final split: 
in the zero-end (Z) treatment both players receive 0, whereas in the fair-end (F) treatment both players receive 50\% of the resource.
The notation $\pi_{i}(t)$ denotes the payoff obtained by Player $i$ if the game ends at Step $t$.}
\label{fig:1}
\end{figure}

In this study, we focus on reciprocal exchanges of resources, modelled through a mixed-motive social dilemma known as the \emph{Incremental Centipede Game} (henceforth ICG)~\citep{rosenthal_games_1981}.
The ICG is a sequential game that involves two individuals who take turns, over a total of $L$ steps\footnote{Each decision step will be denoted by an integer $t\in\left\{1,2,\dots, L\right\}$. We will occasionally use the terms `turn' and `move' to refer to the moments in which each player can act. For $L$ steps, each player has $L/2$ turns.}, to decide about the splitting of a resource (see \autoref{fig:1} for an example).
At each step, the focal player must choose between two actions ($T=$\textit{Take}, $P=$\textit{Pass}).
Playing $T$ on a given step means ending the exchange and receiving a larger share of the resource than the co-player.
Choosing $P$ means passing the decision to the co-player in the following turn (except for the last turn, where the choice involves two different splits).
When the resource to be shared grows with each step of the game, the payoff structure of the ICG typically poses a conflict between the immediate self-interest of one player and the mutual long term benefit of both players~\citep{mckelvey_experimental_1992, palacios-huerta_field_2009}.
This conflict of interest is fuelled by the uncertainty about the goals of the other player, and leaves room for a diverse set of thought processes~\citep{krockow_exploring_2016, lenaerts_evolution_2024, 10.1098.rsif.2025.0153}.

For this experiment, we propose two variants of an ICG (with $L=6$) where the resource grows linearly at each step:
a \emph{zero-end} condition, where the final step is a deadline leading to zero payoffs for both players~\citep{aumann2024irrationality}, and a \emph{fair-end} condition, where the last node leads to each player receiving half of the resource~\citep{brams2020note} (see \autoref{fig:1}).
Although relatively unexplored~\citep{krockow_review} compared to their well-known exponential~\citep{mckelvey_experimental_1992} and constant-sum~\citep{fey_experimental_1996} counterparts, these linear variants combined with different terminal outcomes can capture relevant real-life strategic interactions.
For example, while linear increases keep rewards within plausible ranges, the presence of a deadline at the end of the zero-end game can represent the dangers of waiting too long to harvest a crop, whereas a fair final division can model the successful agreement after an iterative process of negotiations between two parties who distrust each other.

Furthermore, these variations stand on the two opposite sides of a spectrum: 
on the one hand, in the zero-end game, there is no incentive to reach the final decision step, and each choice revolves around anticipating the opponent's move to get the largest share of the resource~\citep{krockow_competitive}; on the other hand, in the fair-end game, inequality-averse and prosocial players might have the intention of reaching the final split, yet they do not have any information about the goal of their opponent. 
This friction makes the decision process in the latter case more nuanced, with motives such as trust, optimism, and risk-taking playing a crucial role in the outcome of the game.
Nevertheless, a classical game-theoretical analysis through backward induction predicts the same subgame-perfect equilibrium for both cases, where the game ends at the first decision step~\citep{ponti_cycles_2000}.
Indeed, in both instances of the ICG, a strictly self-interested and purely rational player should play $T$ at every opportunity, knowing that an equally rational opponent would do the same.

The objective of the present analysis is to go beyond this counter-intuitive prediction and measure the behavioural differences between these two settings, while also uncovering the potential causes behind such differences.
Our primary hypothesis therefore regards the comparison between the outcomes of the two games.
In particular, we expect participants in the fair-end treatment to reach later terminal nodes than the participants in the zero-end treatment, who have less incentives to deviate from the rational strategy.
A secondary hypothesis concerns how the outcome of the first game affects participants’ choices in the subsequent iteration to understand which behavioural patterns are reinforced when the ICG is repeated a second time. 
Specifically, we expect that when a participant manages to stop at the desired extraction point, they make a similar stopping decision in the second round. 
Conversely, when a participant is prevented from stopping at their preferred point in the first round (because their co‐player stopped earlier), the stopping decision is different in the subsequent round.
Finally, through an exploratory analysis, we measure to what extent the behaviour observed in the ICG correlates with certain individual features of the participants, such as prosocial tendencies and cognitive effort.

The exploratory analysis is based on two individual tasks that complement the experimental design, namely the Social Value Orientation task (SVO,~\citep{murphy_measuring_2011}) and the Cognitive Reflection Test (CRT,~\citep{frederick2005cognitive, pennycook2016cognitive}).
The SVO is a resource allocation task that measures how much weight a person gives to the welfare of others in relation to their own.
Participants typically fall into one of four categories: 
(i) competitive, if they maximise their gains and minimise the others'; (ii) individualistic, if they are concerned only with their own gains; (iii) cooperative, if they maximise their outcomes as well as the others'; and (iv) altruistic, if they are willing to sacrifice their own gains in favour of the others.
The CRT is a widely used set of questions designed to assess individuals' ability to suppress an intuitive and spontaneous incorrect answer in favour of a more reflective and deliberative correct answer.
The main assumption underlying this test is the so-called \emph{dual process theory}, which distinguishes between two types of cognitive processing: “System 1”, which is fast, automatic, and intuitive, and “System 2”, which is slower, effortful, and analytical~\citep{kahneman2013perspective, otero2023cognitive}.
These tasks are motivated by the fact that previous attempts to reconcile theoretical and empirical results in the ICG typically rely either on other-regarding preferences, which may distort self-interested decision-making~\citep{bela_altruism_2022, krockow_exploring_2016, gamba_preferences-dependent_2015}, or on bounded rationality, which can undermine backward-induction reasoning~\citep{mckelvey_quantal_1995, mckelvey_quantal_1998, kawagoe_level-k_2012}.
Further details on these two tasks are provided in the Methods (see \autoref{subsec:design}).

The contributions of our work are summarised as follows.
First, we show that the presence of a fair final division significantly increases the likelihood that participants delay their stopping decision in the ICG.
While the distribution of the outcome in the zero-end treatment reaches its peak on the second and third decision steps, the distribution in the fair-end treatment is bimodal, with the two peaks located in correspondence of the fully self-interested and fairness-seeking strategies, respectively.
Thus, a subset of participants appears to prefer the fair outcome over an early gain, even though the former bears more risk and uncertainty.
Furthermore, we show that these behavioural differences can be explained in terms of a minimal evolutionary model fitted on the experimental data, in which different probabilistic strategies compete and propagate via imitation following a replicator equation.
Our theoretical model suggests that, given a setting where the population of players shares a sufficient degree of prosocial motives and noise within the decision-making process, the fair outcome shifts the equilibrium of the system towards a more collectively beneficial strategy.

Secondly, we find that this effect is even more pronounced in the second iteration of the game.
Participants who were prevented from reaching the desired node in the first round undergo an adaptive process in the second round of the game.
The pattern of this adaptation is similar for both treatments, although the probability of changing behaviour is systematically higher in the fair-end treatment.
Then, through our exploratory analysis, we find that the participants who depart from subgame-perfect rationality in the ICG tend to exhibit a stronger prosocial orientation in the SVO. 
However, compared with the zero-end treatment, the fair-end treatment includes a higher share of individualists among the non-rational players, suggesting that the fair outcome may function as a nudge~\citep{montero2025self}.
Lastly, the scores in the CRT do not seem to explain behavioural choices in the ICG, suggesting that the observed non-rational behaviour might be the result of both intuitive and deliberative processes.

Through these findings, the present work corroborates the idea that certain individual traits, such as social preferences, have a crucial impact on human decision-making.
These mechanisms, while deviating from the standard game-theoretical notions of self-interest and common knowledge of self-interest, may enable individuals to achieve better outcomes when facing social dilemmas with their peers.

\section{Results and discussion}
\subsection*{The prospect of a fair final split generates bimodal behaviour.}

\begin{figure}[!ht]
\centering
\resizebox{\textwidth}{!}{%
\includegraphics{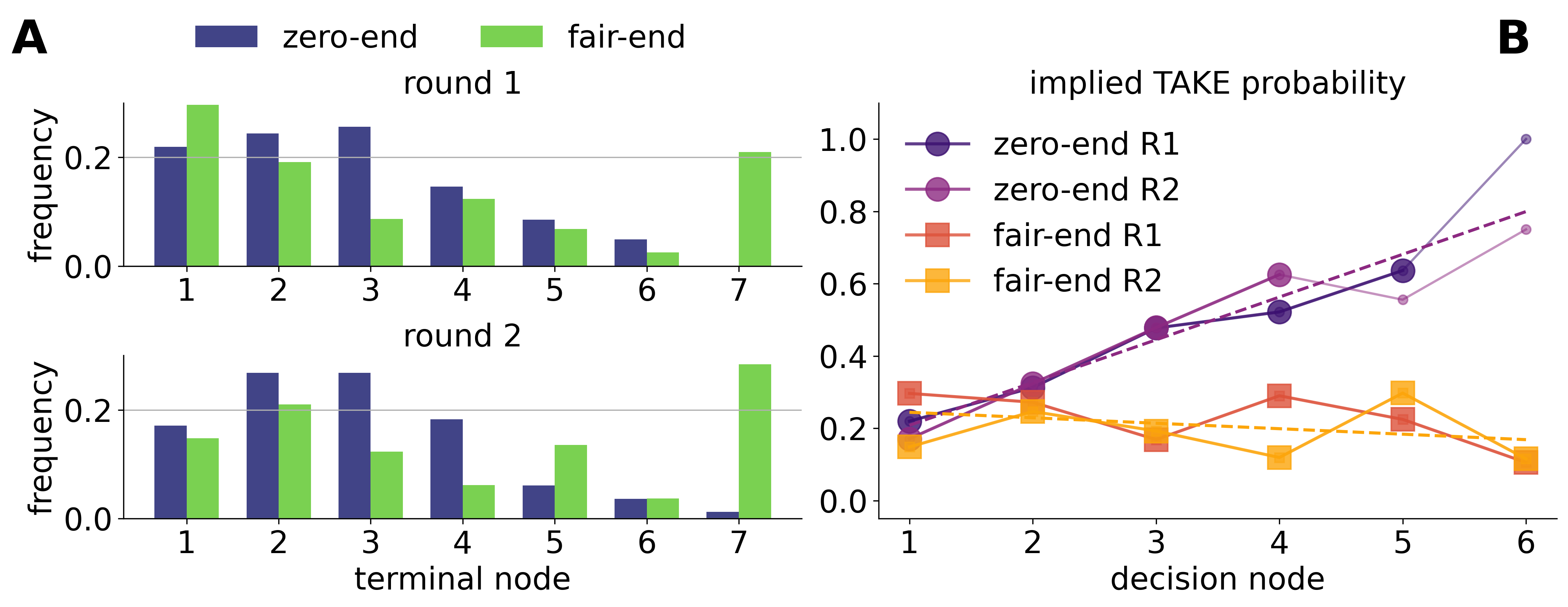}
}
\caption{Outcome of the Incremental Centipede Game.
Panel \textbf{A} reports the distribution of the terminal node for both zero- and fair-end treatment, separately for each round (top: first round; bottom: second round).
The terminal node is the decision step at which one of the two players decides to stop the game, with the convention that the value 7 represents the final possible split. 
Panel \textbf{B} shows the implied probabilities of playing \emph{Take} at each decision step. 
Each point is the proportion of games among those that reached decision step $t$, in which the subject moving at node $t$ chose to \emph{Take}. 
If we define the proportion of games ending at the terminal node $t$ by $f_{t}$ then the implied take probability at node $t$, $p_{t}$, is defined as $p_{t} = f_{t} / \sum_{k\ge t} f_{k}$.
In the zero-end treatment, some markers have smaller size (node 6 for round 1 and nodes 5, 6 for round 2) because the estimate is based on less than 10 samples.
The dashed lines are computed via a simple linear regression separately for each treatment.
Sample sizes: $n_{Z}=82$, $n_{F}=81$ (dyads per round).
}
\label{fig:outcome-cg}
\end{figure}

Let us start by analysing the outcome of the ICG, visualised in \autoref{fig:outcome-cg}. 
In Panel \textbf{A}, we compare the distribution of the terminal node of the games between zero- and fair-end treatments, separately for each round.
The terminal node represents the decision step at which one of the two players decides to conclude the game (i.e. play \emph{Take}), with the convention that the value $L+1=7$ represents the final possible outcome reached when both players always choose to \emph{Pass}.

Regarding the zero-end treatment, the terminal node distribution peaks at the second and third decision step and then rapidly decreases for subsequent nodes.
This distribution is aligned with previous studies~\citep{krockow_competitive, krockow_review}, further confirming the limitations of theoretical solution concepts such as the sub-game perfect equilibrium in this sequential dilemma, even when future stakes are low.
In regard to the fair-end treatment, the distribution is rather bimodal, having its peaks located at the last and first (second) nodes in the first (second) iteration of the game.
In this case, the distribution has greater variability between rounds compared to the zero-end treatment, with a strong decrease in participants who stopped the game on the first decision step (from 30\% to 15\%), and a consequent increase in participants who stopped the game at later nodes in the second round.

In Panel \textbf{B}, we display the implied probability of playing \emph{Take} at each decision step, a measure corresponding to the proportion of games among those who reached the decision step $t$, in which the subject who moves at node $t$ chooses to \emph{Take}.
Formally speaking, if we define the proportion of games that end at the terminal node $t$ by $f_{t}$ then the implied take probability at the node $t$, $p_{t}$, is defined as $p_{t} = f_{t} / \sum_{k\ge t} f_{k}$.

We first notice that the difference between the two distributions, compared with Panel \textbf{A}, is further highlighted.
Although the two treatments share similar probability values for the first two decision steps, they drastically diverge for subsequent nodes.
In the zero-end treatment, where the last outcome does not bring any benefit to either participant, the implied take probability increases linearly (with a slope coefficient $\beta = 0.12$ estimated via linear regression, see dotted lines). 
In contrast, the presence of the fair outcome at the end of the second treatment keeps the implied take probabilities bounded at an approximately constant value of 20\% (indeed, in this case, the estimated slope coefficient decreases to $\beta = -0.02$).

Our primary hypothesis that the distribution in the fair-end treatment is higher on average than in the zero-end is confirmed by a one-sided Wilcoxon rank sum test (see \autoref{subsec:statistical-models}) applied to the average of the terminal node distributions over the two rounds ($W = 16379$, $p < 0.001$, $n_{Z}=164$, $n_{F}=159$, Hodges–Lehmann location shift $\hat{\Delta}=1$, 95\% one-sided lower CI $\Delta\ge0.5$).
Additionally, to better quantify the effect of different final splits on the outcome of the game, we fit a cumulative link model with partial proportional-odds (PPOM, see \autoref{subsec:statistical-models}) based on seven predictors:
the treatment (0 if zero-end, 1 if fair-end), the round number, the age and the sex of both participants in the dyad, as well as the binary variable indicating whether the dyad is created according to perfect stranger matching.
The proportional odds assumption is removed for the treatment variable.

A likelihood ratio test between the PPOM, whose parameters are reported in bold in \autoref{tab:ppom} in \autoref{appendix:tables}, and the reduced model that does not include the treatment variable confirms the importance of the impact of the treatment on the behavioural choices of our participants:
The former, in fact, is a significantly better fit than the reduced model (LR, $p < 0.001$; AIC $1133.8$ vs $1185.4$).
Overall, our model indicates that, relative to the zero-end treatment, the fair-end treatment shifts probability mass away from intermediate nodes and mostly toward the final node.
Specifically, the average marginal effects of the treatment variable, i.e. the sample-average change in the predicted probability of each ordinal category for a one-unit increase in that predictor, show that shifting from the zero-end to the fair-end treatment increases the probability of terminal nodes 1, 5, and 7 by 1.9, 3.2, and 24.2 pp, respectively. 
Correspondingly, the probabilities of categories 2, 3, 4, and 6 decrease by a combined total of nearly 30 pp (see \autoref{tab:ame} in \autoref{appendix:tables}).

\subsection*{Behavioural differences between treatments are explained by an evolutionary model.}

To explain the behavioural differences discussed previously, we propose a minimal model based on evolutionary game theory~\citep{smith_logic_1973,smith_evolution_1982,sigmund_calculus_2010,fernandez_domingos_egttools_2023}.
The model, described in detail in the Methods (\autoref{subsec:dynamics}), combines probabilistic strategies and social preferences to study the evolution of behavioural types in a population of interacting agents.
We propose five different types of players, each equipped with a probabilistic strategy that specifies the probability of choosing \emph{Take} at any turn $t$ (given that such a turn $t$ is reached), denoted $P(\text{Take} \mid t)$.
We consider three types for which the \emph{Take} probability remains constant over every turn of the game: (1) a constantly low ($CL$) \emph{Take} probability, $P(\text{Take} \mid t)=p<0.5$, (2) a constantly medium ($CM$) \emph{Take} probability, $P(\text{Take} \mid t)=0.5$, and (3) a constantly high ($CH$) \emph{Take} probability, $P(\text{Take} \mid t)=1-p>0.5$.
Regarding the remaining two types, we consider a variable \emph{Take} probability: (4) a linearly increasing ($LI$) \emph{Take} probability from the lower to the higher value (i.e, $p\rightarrow0.5\rightarrow1-p$), and (5) the linearly decreasing ($LD$) counterpart (i.e, $1-p\rightarrow0.5\rightarrow p$).
As we can see, the strategies proposed in our model are stochastic and modulated by a free parameter $p\in[0, 0.5)$.

Our model then incorporates social preferences by defining the payoff of player $A/B$, $\Pi_{A/B}$, as the weighted sum of one's own expected material payoff, $E[\pi_{A/B}]$, and the other player's material payoff, $E[\pi_{B/A}]$. 
In formulas, the payoffs are therefore defined as $\Pi_{A/B} = E[\pi_{A/B}] + \delta E[\pi_{B/A}]$, where
the second free parameter of our model, $\delta\in[-1,+1]$, specifies the weight and the orientation of the social preferences of the population.
Each pair of values $(p,\ \delta)\in [0, 0.5)\times[-1,+1]$ produces different equilibria and, consequently, different terminal node distributions.
Therefore, we fit the model on our experimental data, finding two optimal parameter configurations: $p=0.25$ and $\delta=0.1$ for the zero-end treatment, and $p=0.21$ and $\delta=0.4$ for the fair-end treatment (details in \autoref{subsec:dynamics}).
Notably, while the optimal level of noise in the strategies, $p$, remains similar across treatments, the prosocial orientation $\delta$ is considerably higher in the fair-end game, underscoring the importance of social preferences in explaining human behaviour in this particular mixed-motive setting.

\begin{figure}[!ht]
\centering
\resizebox{\textwidth}{!}{%
\includegraphics{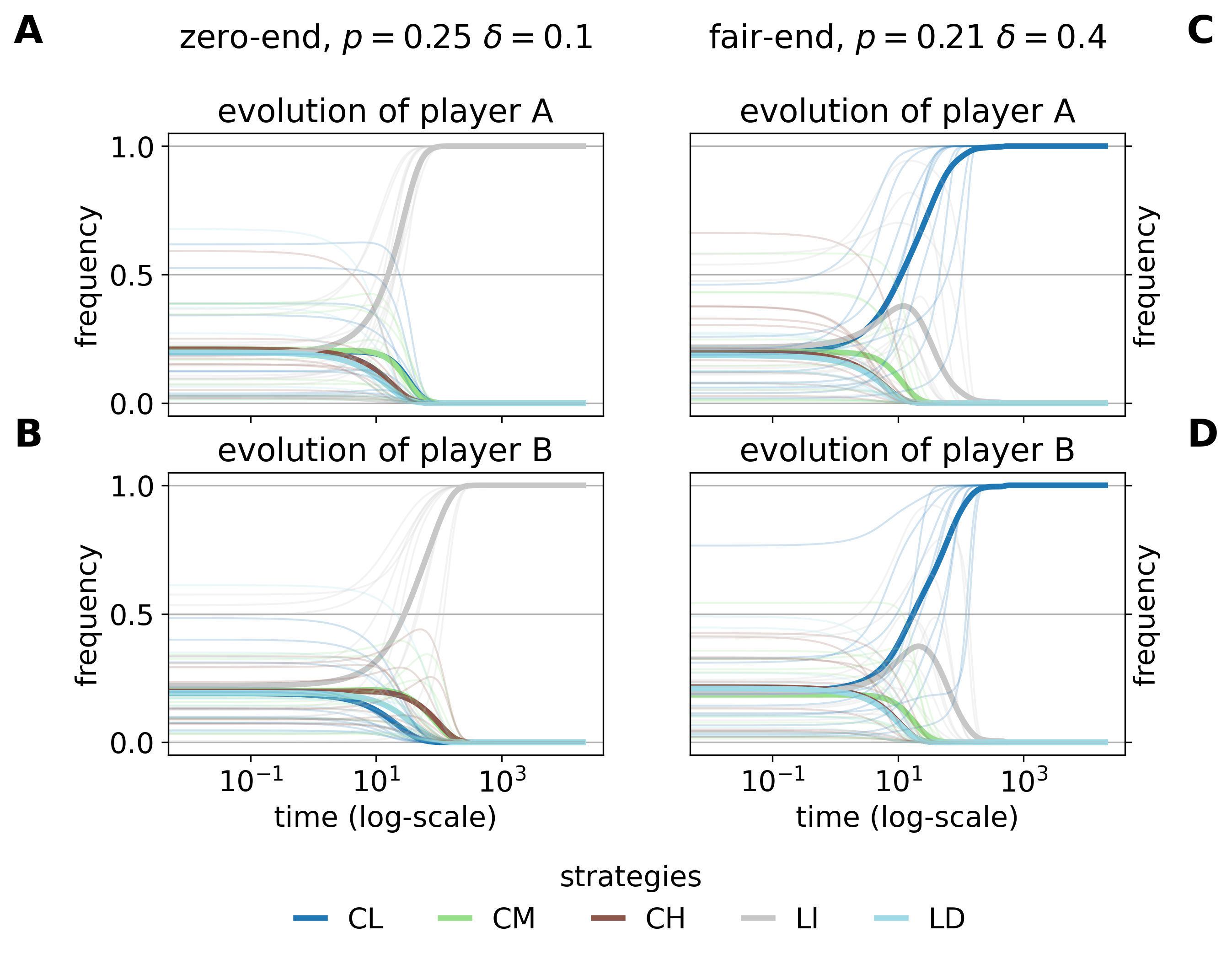}
}
\caption{Evolutionary dynamics in the Incremental Centipede Game. 
Each panel shows how the frequency of the strategies of our model evolves over time according to an asymmetric replicator dynamics, separately for each role and treatment.
Panels \textbf{A,B} and \textbf{C,D} refer to the zero- and fair-end treatments, respectively.
The top panels (\textbf{A,C}) show the evolution of the population of players in role A, while the bottom panels (\textbf{B,D}) represent the players in role B.
The obtained frequencies are the average of 200 deterministic processes starting from random initial points, with a time horizon $\tau=2\cdot10^4$.
The thinner, dimmed lines show a sample of 10 trajectories of the dynamical system.
The strategies define different \emph{Take} probabilities for each turn of the player, and follows the notation in the manuscript: $CL:=$ constantly low \emph{Take} probability ($p$), $CM:=$ constantly medium \emph{Take} probability ($0.5$), $CH:=$ constantly high \emph{Take} probability ($1-p$), $LI:=$ linearly increasing \emph{Take} probability ($p\rightarrow0.5\rightarrow 1-p$), $LD:=$ linearly decreasing \emph{Take} probability ($1-p\rightarrow0.5\rightarrow p$).
In both treatments, the values of the parameters $(p,\delta)$ are based on the fitting with our experimental data.
}
\label{fig:replicator}
\end{figure}

The evolution of the aforementioned types follows an asymmetric replicator equation~\citep{Accinelli2011} (see \autoref{subsec:dynamics}), whose results are visualised in \autoref{fig:replicator}. 
In this deterministic process, the portion of the population with a strategy that yields a higher-than-average expected payoff is imitated by others and propagates within the population.
Each panel shows the evolution of each strategy over time ($0\leq t \leq 2\cdot10^4$), separately for each treatment and player's role.
Panels \textbf{A-B} on the left refer to the zero-end treatment, whereas Panels \textbf{C-D} on the right refer to the fair-end treatment.
The top (bottom) panels depict the evolution of the population of individuals in role A (B). 
The obtained frequencies are the average of 200 trajectories starting from random initial configurations of the population.

As we can see, the dynamical system closely reproduces the results presented in \autoref{fig:outcome-cg}.
Indeed, in the zero-end treatment, the dynamical system converges to a state where the whole population chooses \emph{Take} with a linearly increasing ($LI$) probability in both roles, which leads to a bell-shaped terminal node distribution.
The presence of a fair final outcome, instead, causes both populations to shift to a constantly low ($CL$) \emph{Take} probability, hence producing a bimodal terminal node distribution.
Our model therefore suggests that, given a setting where the population of players shares a certain degree of prosocial motives and noise within the decision-making process, the fair division at the end of the ICG shifts the equilibrium of the system to a more collectively beneficial outcome.

\subsection*{Players prevented from reaching the desired outcome in the first game adapt to achieve it in the next one.}

After studying the outcome of the ICG at the dyadic level, we now move to a behavioural analysis at the individual level.
Specifically, our goal is to understand how the outcome of the first iteration of the ICG influences participants’ behavioural choices in the subsequent round.
Accordingly, this analysis aims to determine under which conditions players are more likely to change their strategy between rounds.
A strategy profile corresponds to the number of \emph{Pass} decisions (an integer between 0 and 3) that ultimately leads the player to their preferred split - unless they are anticipated by an earlier \emph{Take} move of their co-player.

\begin{figure}[!ht]
\centering
\resizebox{\textwidth}{!}{%
\includegraphics{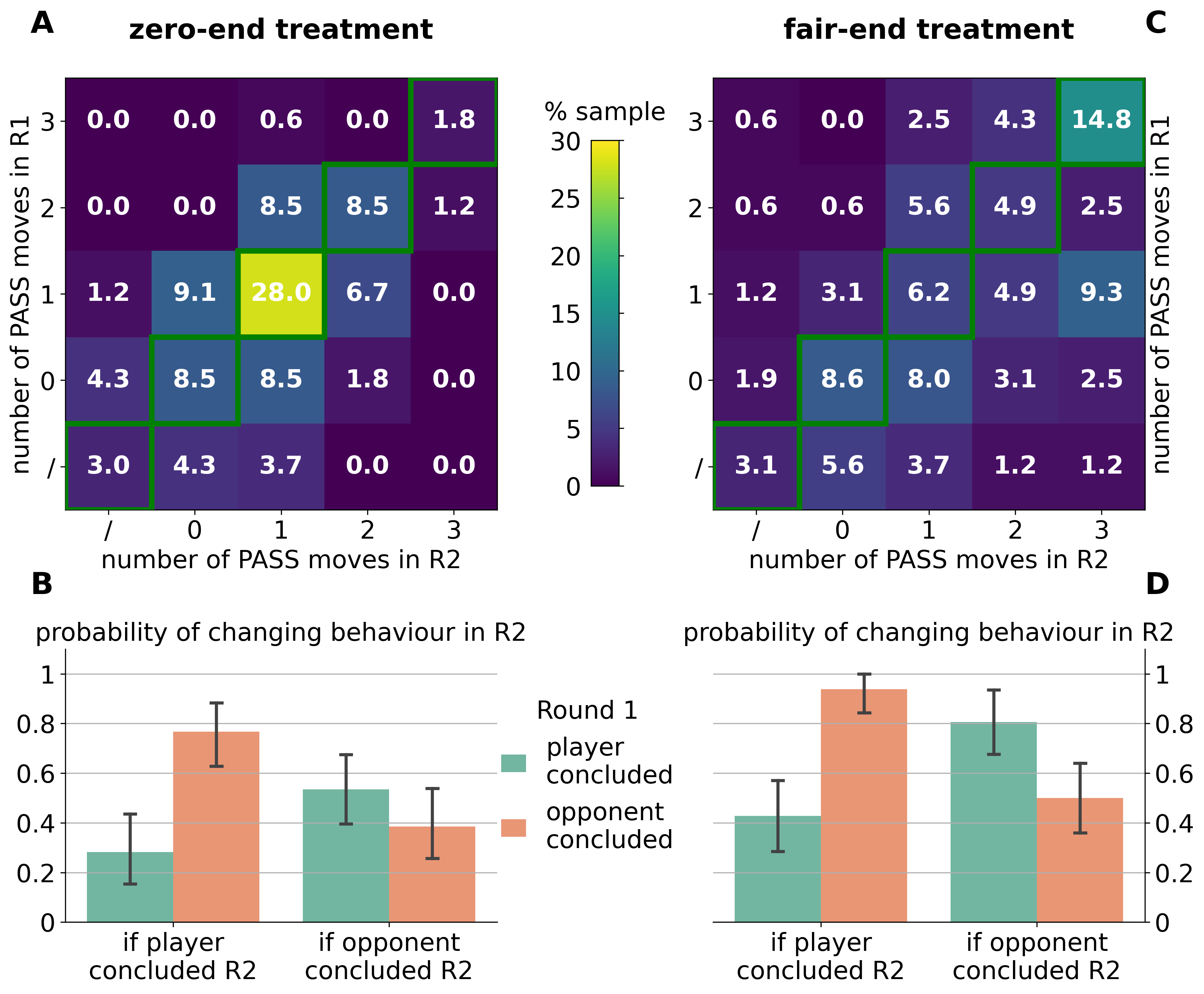}
}
\caption{Behavioural changes between the first and the second iteration of the Incremental Centipede Game.
The heat maps in the top panels (\textbf{A} and \textbf{C} for the zero- and the fair-end treatments, respectively) contain the sample's percentage for each possible strategy profile in both rounds of the game. 
Each cell $(i,j)$ with $i,j\in\left\{0,1,2,3\right\}$ contains the percentage of players with $i$ number of \emph{Pass} decisions in round 1 and $j$ number of \emph{Pass} decisions in round 2.
The label `/' represents the players in role B facing an opponent A who stopped the game at the first decision step.
The diagonal entries (highlighted in green) represents the participants who made the same number of \emph{Pass} decisions in both rounds.  
The bottom panels (\textbf{B} and \textbf{D} for the zero- and the fair-end treatments, respectively) show the estimated probability of changing behaviour as a function of whether the player took the last decision of the game or the opponent did.
Each panel contains two categories, namely the players who did and did not reach the desired terminal node in the second round.
The black vertical bars represent the 95\% confidence intervals.
Sample sizes: $n_{Z}=164$, $n_{F}=162$ (individuals).
}
\label{fig:r1r2}
\end{figure}

Here, we focus on a binary classification of behaviour, i.e., whether a participant’s number of \emph{Pass} decisions in the second round is equal to or different from that in the first round.
As stated in the Introduction, we hypothesise that one of the primary drivers of behavioural changes is whether a participant was prevented by their co-player from stopping at their preferred extraction point in the first round.
This hypothesis is supported by our experimental data, as illustrated in \autoref{fig:r1r2}.

First, the heat maps in Panels \textbf{A} and \textbf{C} represent the percentage of our sample for each pair of possible choices in the two rounds, separately for each treatment.
In both rounds, strategic choices are characterised by the number of turns the player chooses the \emph{Pass} option.
The entries labelled with `/' in the figure represent the scenarios in which a player in role B faces an opponent who stops the game at the first decision step, thus providing no information about such a player.

In the zero-end treatment (Panel \textbf{A}), the sample is balanced and almost symmetrically distributed between players who followed the same strategy in both rounds (diagonal entries, highlighted in green) and players who reached later/earlier nodes in the second iteration of the game (lower-triangular/upper-triangular entries, respectively).
Indeed, approximately half of the participants did not modify their behaviour.
In particular, the peak of the distribution ($\sim28\%$ of the sample) is associated with players who chose to \emph{Pass} once in both rounds.
Then 26\% made a higher number of \emph{Pass} decisions in the second round than in the first, and the remaining 24\% of the sample made a lower number of \emph{Pass} decisions in the second round than in the first.
Moreover, the entries furthest from the diagonal account for a negligible portion of the sample.

Interestingly, the behavioural landscape changes in the fair-end treatment (Panel \textbf{C}), leading to a more asymmetrical distribution.
Consistent with the findings shown in \autoref{fig:outcome-cg}\textbf{A}, participants who increased the number of \emph{Pass} decisions cover 42\% of the sample, followed by 38\% who made the same number of \emph{Pass} decisions in both rounds. 
The peak of the distribution lies in the latter group, where approximately 15\% of the participants chose to reach the final split in both rounds.
Lastly, only 20\% of the sample decreased the number of \emph{Pass} decisions.

This pattern indicates that, in the fair-end treatment, participants adapted their strategy in favour of reaching the socially optimal outcome in the second iteration of the game.
This shift towards mutual cooperation may arise because any \emph{Pass} move experienced in the previous round inherently contains a signal for cooperation.
Participants may consequently use these cues to create a mental representation of their future co-player and to update their beliefs on the other's potential strategy.
This belief update can, in turn, bias the individual's decision-making towards a more optimistic or risk-taking attitude, as suggested by previous studies~\citep{lenaerts_evolution_2024, 10.1098.rsif.2025.0153}.

Panels \textbf{B} and \textbf{D} illustrate the link between the estimated probability of changing behaviour and the two binary variables (one for each iteration of the game) that indicate whether the player reached the desired terminal node of the game or was prevented from reaching it by their co-player.
As we can see, both treatments share the same ordering, although the values are systematically higher in the fair-end treatment.
Among the subset of participants who managed to reach the preferred outcome in the second repetition of the game, the individuals who were prevented from reaching it in the first iteration show the highest rate of behavioural change.
In contrast, the participants who made the last decision in both games have the lowest rate of change.
The opposite happens, even though not as marked, when we look at the participants who did not manage to reach the preferred outcome in the second round.

To explore in more depth the sample of participants who implemented their final decision in the second round, we fit a logistic model, and its summary is reported in \autoref{tab:model2}, under \autoref{appendix:tables}.
The model, separated by treatment, predicts the binary target variable of behavioural change using five predictors associated with each participant:
the participant's age and sex, their role in the game, whether the participant met a stranger in the second round, and whether the participant was able to end the game at their preferred extraction point in the first round.
In both treatments, the only significant predictor appears to be the latter ($p<0.001$ for both treatments).
Specifically, ending the first round systematically decreases the odds of changing behaviour when holding all the other predictors constant (odds-ratio: $OR=0.11$, 95\% CI $[0.04, 0.31]$, in the zero-end treatment and $OR=0.05$, 95\% CI $[0.01, 0.20]$, in the fair-end treatment).

\subsection*{Prosocial players tend to reach later terminal nodes.}

\begin{figure}[!ht]
\centering
\resizebox{\textwidth}{!}{%
\includegraphics{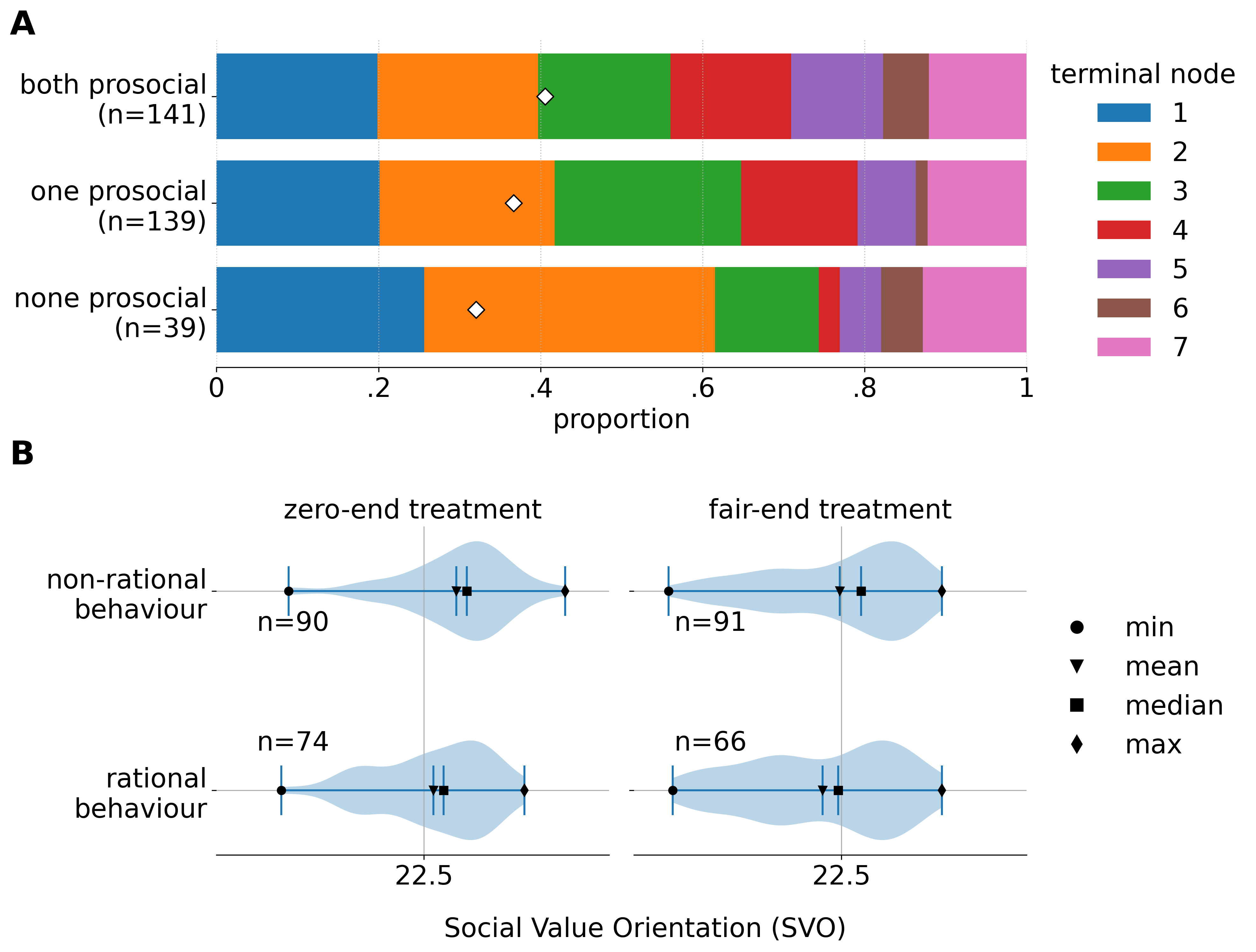}
}
\caption{Relation between participants' social value orientation (SVO) and the terminal node of the Incremental Centipede Game.
Panel \textbf{A} (top) shows the distribution of the terminal node for three distinct classes of dyads, namely dyads with zero, one, or two prosocial players.
The white markers denote the mean of each distribution (scaled between 0 and 1).
In Panel \textbf{B} (bottom), the violin plots show the distribution of the SVO score of the participants who followed the rational strategy (i.e., number of \emph{Pass} decisions $=0$) at the bottom, while the top distributions are associated to the participants who followed a non-rational strategy (i.e., number of \emph{Pass} decisions $>0$).
The value 22.5 on the horizontal axis represents the boundary between individualistic (SVO$^{\circ}<22.5$) and prosocial (SVO$^{\circ}>22.5$) orientation.
The results are visualised separately for both treatments. 
}
\label{fig:svo-cg-explo}
\end{figure}

The results obtained with our theoretical model suggest that social preferences might play an important role in strategic decision-making in the context of the ICG.
To further examine this relationship, we conducted an exploratory analysis linking the collective outcomes observed in the ICGs to participants' social preferences.
As mentioned in the Introduction, this analysis is based on the SVO task, which assigns each participant an individual score.
The SVO score quantifies the degree to which individuals value others’ gains relative to their own on a continuous scale.
The findings of the exploratory analysis are presented in \autoref{fig:svo-cg-explo}.

Panel \textbf{A} shows how the distribution of the dyad outcome, i.e., the terminal node of the game, varies as a function of the number of prosocial participants in the dyad (i.e., whether a dyad is composed of zero, one or two players with an SVO score greater than $22.5^{\circ}$).
Overall, selfish behaviour declines as the number of prosocial players in the dyad increases.
Correspondingly, the mean of each distribution slightly increases with the number of prosocial members in a dyad (from 2.92 to 3.20 and finally 3.43).
The most pronounced distribution shifts occur between B-players stopping at the second node (36\% when the dyads have no prosocial player versus $\approx20\%$ when the dyads have at least one prosocial player), as well as B-players stopping at the fourth node (3\% when the dyads have no prosocial player versus $\approx14\%$ when the dyads have at least one prosocial player).
These results suggest that the effect of prosocial motives on the decisions made in the ICG may depend on the role assumed by the player, with a more prominent effect for players in role B. 

Panel \textbf{B} focuses on the inverse relationship, that is, how the SVO distribution of the players shifts as a function of their individual choices in the ICG.
In particular, the violin plots show, separately for each treatment, the distribution of SVO scores in two subgroups of our sample: 
rational (zero \emph{Pass} moves) and non-rational (at least one \emph{Pass} move) participants. 
In both treatments, rational participants exhibit a heavier left tail than non-rational participants, indicating a stronger presence of individualists among the rational players. 
This result supports the close association between rational behaviour and self-interest.
Notably, however, in the fair-end treatment, the non-rational subgroup still shows a pronounced left tail, whereas this feature is largely absent in the zero-end treatment.
This result suggests that the prospect of a fair future outcome at the end of the ICG might act as a nudge~\citep{montero2025self}, i.e., it can potentially convince individualists to pursue a more cooperative strategy.

A one-sided Wilcoxon rank sum test applied separately to each treatment confirms that the score distribution differs significantly between these subgroups in both treatments.
(zero-end: $W = 2512$, $p < 0.05$, $n_{\text{rational}}=74$, $n_{\text{non-rational}}=90$, $\hat{\Delta}=-0.076\text{ rad}$, 95\% one-sided lower CI $\Delta\le-0.034 \text{ rad}$;
fair-end: $W = 2522$, $p < 0.05$, $n_{\text{rational}}=66$, $n_{\text{non-rational}}=91$, $\hat{\Delta}=-0.048\text{ rad}$, 95\% one-sided lower CI $\Delta\le-3.696e-05\text{ rad}$.)
Furthermore, to assess the extent to which players' SVO affects the outcome of the ICG, we extend the PPOM reported in \autoref{tab:ppom} in \autoref{appendix:tables} by adding two predictors, namely the SVO score of both players in each dyad.
The coefficients of the updated model are reported in \autoref{tab:model3} in \autoref{appendix:tables}. 
Incorporating these variables significantly improves the fit of the model (LR, $p<0.01$; AIC 1125.8 vs 1133.8).
Both predictors are statistically significant, with higher SVO values increasing the chances of reaching later terminal nodes (odds-ratio: $OR\approx1.25$, 95\% CI: $(1.02, 1.54)$, for the SVO of player A and $OR\approx1.33$, 95\% CI: $(1.08, 1.64)$, for Player B).

While our exploratory analysis suggests that social preferences have a significant influence on decision-making in the ICG, the same cannot be concluded for cognitive reflection as measured by the CRT.
Participants' performance on the CRT, measured through the percentage of correct and intuitive answers, does not show a significant correlation with the outcome reached in the ICG.
This result is consistent with previous studies~\citep{baghestanian2016go}, arguing that deviations from game-theoretical rationality observed in the ICG are unlikely to be caused by limited cognitive effort.
Additionally, previous studies on the CRT suggest that, while the original version of the CRT has been widely validated \citep{li2024does,branas2019cognitive}, the overexposure to the CRT increases the chances of familiarity among participants, thus becoming a major concern regarding its viability \citep{thomson2016investigating, stieger2016limitation}.
In our case, the model summarised in \autoref{tab:crt} of \autoref{appendix:tables} suggests that a high score in the CRT slightly increases the likelihood of reaching later terminal nodes.
However, the estimated coefficients are not significant, and more experiments are needed to prove whether there exists a link between cognitive reflection and strategic decision-making in social dilemmas.

\section{Conclusions}

Our behavioural experiment investigated the impact of a fair but uncertain outcome on strategic decision-making during reciprocal exchanges of resources.
Our findings suggest that certain individual traits, such as social preferences, allow humans to overcome self-interest and achieve better collective outcomes.
Indeed, in line with the existing literature~\citep{krockow_review}, the participants substantially deviated from the purely self-interested and rational behaviour in both treatments of our experiment.
Particularly in the fair-end treatment, a large portion of the sample displayed a fairness-seeking cooperative strategy to reach the fair division in the final step of the game.
We then stressed the importance of other-regarding motives in this context, both theoretically and experimentally.
First, our evolutionary model showed that if a population of players is sufficiently prosocial and follows probabilistic strategies, the fair split causes a shift towards a mutually beneficial strategy.
Secondly, the experimental data validated these theoretical findings, as participants who departed from self-interested behaviour exhibited a stronger prosocial orientation in the SVO. 
Finally, the sharp increase in fair outcomes in the second repetition of the ICG suggests that the cooperative signals received in the first round might have encouraged risk-averse prosocial individuals to engage in cooperative behaviour.
Additionally, features associated with hot cognition~\citep{raggioli2025computational}, such as empathy and guilt, might have potentially influenced and shifted the subsequent choice of non-cooperative individuals.
Future research should further address this link between altruism and empathy, as it can constitute a necessary ingredient to shift from the classical paradigm of self-interest and expectation of others' self-interest to a more flexible framework, in which individuals have diverse social motives and form potentially different beliefs about the nature of their peers.

\section{Methods}
\subsection{Experimental design}\label{subsec:design}

We conducted a between-subjects behavioural experiment with two treatments.
The experiment, which consists of three tasks, proceeded as follows.
After accepting the informed consent form, participants receive all the necessary information to complete the experiment, including instructions and payment details. 
The participants then begin the session with the Social Value Orientation (SVO) task~\citep{murphy_measuring_2011, montero2025self}. 
This individual task consists of six decisions on how to share a sum of money with another person.
The participants then proceed to the second task, i.e., the Incremental Centipede Game (ICG). 
If they successfully complete a comprehension test about the ICG, they are randomly assigned a role (Player A or B) that will remain fixed throughout the session. 
They are then placed in groups of four people and play the ICG for two rounds, each time with a different participant randomly selected from the group. 
The only difference between the two treatments lies in the type of ICG played by the participants:
In the zero-end treatment, the final node constitutes a deadline (i.e., zero payoffs for both players)~\citep{aumann2024irrationality}, while in the \emph{fair-end} treatment the final node leads to a 50-50 split of the resource~\citep{brams2020note} (see \autoref{fig:1}).
The third and final task is the Cognitive Reflection Test (CRT)~\citep{frederick2005cognitive}, which consists of a set of four short numerical problems.
Finally, a post-experiment survey is presented to the participants, aimed at qualitatively assessing personal characteristics such as trust, risk aversion, familiarity with the tasks, as well as prior knowledge of game theory and behavioural experiments (see \autoref{appendix:survey}).

The payoffs (computed in Experimental Currency Units, or ECUs) are converted to a financial bonus in GBP (\textsterling)\footnote{In our experiment \textsterling1.00 =100 ECUs.} according to the following conditions.
Regarding the SVO, each participant has a 10\% chance of being selected to receive a financial bonus based on their choices.
The selected participants are randomly assigned one of two roles: ``Giver" or ``Receiver".
Givers receive the monetary amount they allocated to themselves in one random sample of their six choices.
The Receivers instead get the amount that another randomly selected participant decided to allocate to the other person.
Concerning the ICG, for each participant, we randomly sample the outcome of one of the two rounds to determine their bonus.
Finally, in the CRT, participants receive 5 ECUs per correct answer.
Note that these financial bonuses are added to a fixed participation fee of 3\textsterling.
To avoid cross-contamination, the payoffs of each task are only announced at the end of the experimental session.
We are nevertheless aware that eliciting the SVO prior to the ICG might potentially influence strategic behaviour in the ICG. 
However, the order in which the tasks are performed remains fixed for all participants, hence making any potential priming effects consistent between treatments and thus unlikely to confound their comparisons.
All the instructions of the experiment can be found in \autoref{appendix:experiment-material}.

This experiment was approved by the ethical Committee of the \textbf{MASKED} on 20/02/2024 and was preregistered in the Open Science Framework (OSF) (\href{https://osf.io/ht56e/overview?view_only=66c7cb5fbdd74a41afc210bf503794b1}{anonymous link}).
Deviations from the preregistration are reported in \autoref{appendix:deviations}.
All experimental sessions were conducted in accordance with the General Data Protection Regulation (GDPR) and followed the practice and protocols traditionally followed in experimental economics.
We ran six and seven sessions for zero- and fair-end treatment, respectively.
The experimental sessions took place regularly from 25/06/2025 until 11/07/2025 and provided on average 25 samples per session.
Participants were recruited online through Prolific, a crowd-sourcing platform, and selected according to the following filters: (i) first language: English; (ii) approval rate: $\geq90\%$; (iii) number of previous submissions: between 10 and 500; (iv) balanced gender distribution.
More information on the demographics of the participants can be found in \autoref{appendix:demographics}.

\subsubsection*{Social Value Orientation (SVO)}
\begin{figure}[!ht]
\centering
\resizebox{\textwidth}{!}{%
\includegraphics{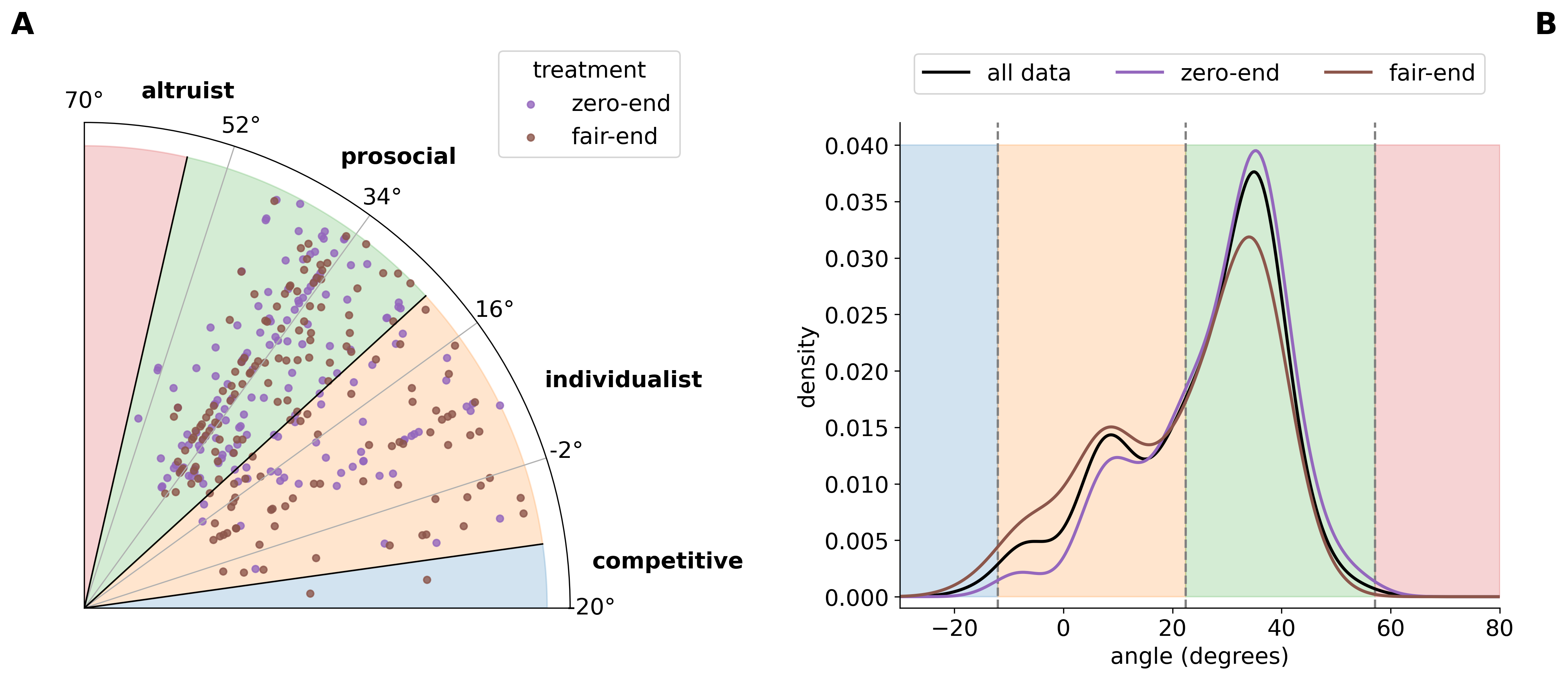}
}
\caption{Distribution of the Social Value Orientation (SVO) of our experimental sample.
Panel \textbf{A} shows the scatter plot in polar coordinates where each individual is represented by a point whose angle $-20^{\circ} \leq\theta\leq 70^{\circ}$ is determined by the individual's SVO and the radius $r$ is randomly assigned to reduce overlaps.
The points are coloured by treatment.
Panel \textbf{B} reports the density function of the SVO distribution (estimated with a Gaussian kernel) for both treatments, as well as for the aggregated dataset.
Both Panels are divided in four regions, namely competitive, individualistic, prosocial and altruistic individuals, according to the boundaries specified in~\citep{murphy_measuring_2011}.
Sample sizes: $n_{Z}=164$, $n_{F}=162$ (individuals).
}
\label{fig:svo}
\end{figure}

Every item in the SVO task has the same general form. 
Each item represents a choice of resource allocation between oneself and another individual over a well defined set of joint payoffs. 
These choices are used to compute the Slider measure~\citep{murphy_measuring_2011}. 
If we label the mean allocation for oneself as $\bar{A}_{s}$ and the mean allocation for the other participant as $\bar{A}_{o}$, the SVO index is computed using the following formula:
\begin{equation}
    SVO^{\circ} = \arctan\left((\bar{A}_{o} - 50) / (\bar{A}_{s} - 50) \right).\nonumber
\end{equation}
Given the original set of rewards proposed in \citep{murphy_measuring_2011}, altruists have an angle greater than 57.15\textdegree; cooperative participants lie between 22.45\textdegree and 57.15\textdegree; individualists lie between –12.04\textdegree and 22.45\textdegree; and competitive participants have an angle lower than –12.04\textdegree. 
The distribution of the SVO of our experimental sample is shown in \autoref{fig:svo}.

\subsubsection*{Incremental Centipede Game (ICG)}
The ICG is a 2-player sequential task. 
Several payoffs' variations exist in the literature (see~\citep{krockow_review} for a review), and each of them entails a certain degree of conflict between shared (cooperative) and opposing (competitive) interests.
In this experiment, the presence of a deadline at the end of the zero-end ICG promotes the competitive tendency of anticipating the opponent. 
The fair final outcome, instead, encourages prosocial players to engage in further exchanges to signal their intention of reaching the final node.
If the final decision step is not reached, the sequential nature of the ICG implies that the game only provides full information about the player who implements their full strategy to achieve the desired split.
Indeed, the dyadic outcome (i.e., the players' joint payoffs) is determined by the terminal node of the game, which is in turn decided by the first player who chooses to stop the ICG.

\subsubsection*{Cognitive Reflection Test (CRT)}
\begin{figure}[!ht]
\centering
\resizebox{\textwidth}{!}{%
\includegraphics{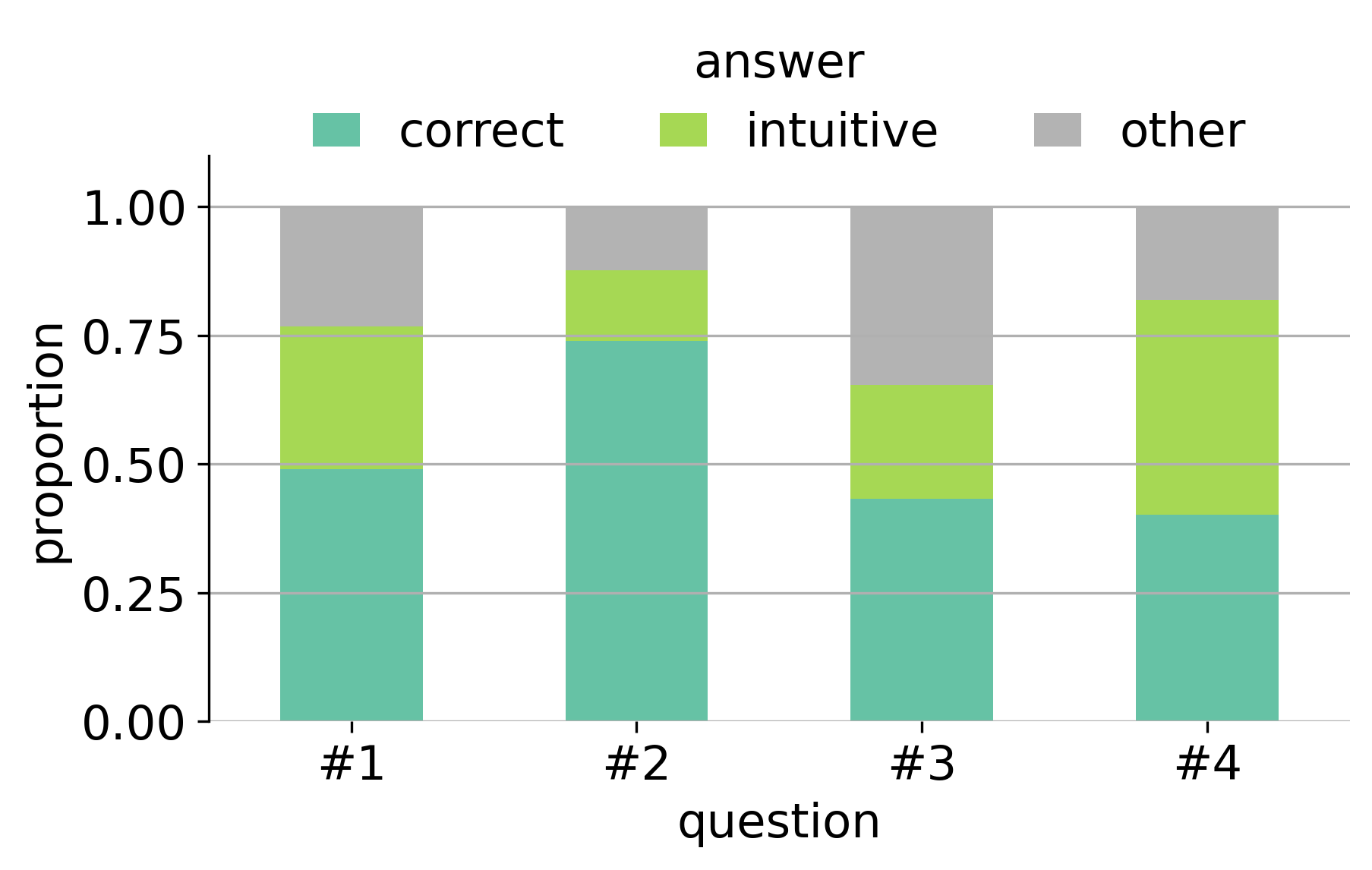}
}
\caption{Results of the Cognitive Reflection Test (CRT).
The stacked bar plot reports the proportion of types of answers given by our experimental sample for each question of the CRT.
Sample size $n=326$.
}
\label{fig:crt}
\end{figure}

The CRT~\citep{frederick2005cognitive, otero2023cognitive} is a set of short numerical problems designed to elicit a spontaneous but incorrect response, requiring further reasoning to arrive at the correct answer.
Although the original version of the CRT has been widely validated~\citep{li2024does,branas2019cognitive}, its widespread use has led to concerns about exposure~\citep{thomson2016investigating, stieger2016limitation}.
To address this issue, alternative variations of the original questions have been proposed~\citep{primi2016development,toplak2014assessing,thomson2016investigating}.
In our implementation, we chose and adapted four questions from the literature. 
Each question has a one-minute timer before moving on to the next.
For each participant, we measure the total number of correct answers as well as the number of intuitive ones.
The selected questions, with the associated intuitive and correct answers, are reported below, while the results of the test are reported in \autoref{fig:crt}.
\begin{enumerate}
    \item In an athletics team, tall members win three times more medals than short members. This year the team has won 60 medals. How many of these have been won by short athletes? (intuitive: 20; correct: 15)
    \item If three elves can wrap three toys in an hour, how many elves are needed to wrap six toys in two hours? (intuitive: 6; correct: 3)
    \item Jerry received both the 12th highest and the 12th lowest mark in the class. How many students are there in the class? (intuitive: 24; correct: 23) 
    \item A cheese and crackers snack costs $2.20$\$ in total. The cheese costs $2.00$\$ more than the crackers. How much do the crackers cost (in cents)? (intuitive: 20; correct: 10)
\end{enumerate}

\subsection{Game-theoretical model}\label{subsec:dynamics}
\subsubsection*{Strategies}
We assume that our population of players consists of five possible types.
Each type is determined by the probability of choosing \emph{Take} during their turns.
Given that the players have three turns each, each strategy can be described by a vector with three elements, $\sigma = [\sigma_{1}, \sigma_{2}, \sigma_{3}]$, where $0\leq\sigma_{i}\leq1$ gives the probability of choosing \emph{Take} at the $i$-th turn.
Our first type is a random player with $\sigma = [0.5,\ 0.5,\ 0.5]$. At each turn, this type of player decides to play \emph{Take} according to a coin toss.
The remaining four types depend on a free parameter $p\in[0, 0.5)$.
Specifically, we define two opposite types of player, one whose \emph{Take} probability is relatively low and fixed at a value $p$ (i.e., $\sigma = [p,\ p,\ p]$) and its counterpart with a relatively high fixed probability $1-p$ (i.e., $\sigma = [1-p,\ 1-p,\ 1-p]$). Lastly, we define two types of players with a variable \emph{Take} probability, one linearly increasing from $p$ to $1-p$ (i.e., $\sigma = [p,\ 0.5,\ 1-p]$), and the other linearly decreasing (i.e., $\sigma = [1-p,\ 0.5,\ p]$).

\subsubsection*{Payoffs}
Let us denote the strategy of Player A and B by $\sigma_{A}$ and $\sigma_{B}$, respectively.
These strategies produce a probability distribution $T$ over the possible outcomes of the game, according to the following formula:
\begin{align*}
    & P(T=1) = \sigma_{A1}\\
    & P(T=2) = (1 - \sigma_{A1})\sigma_{B1}\\
    & \ldots\\
    & P(T=6) = (1 - \sigma_{A1})(1 - \sigma_{B1})(1 - \sigma_{A2})(1 - \sigma_{B2})(1 - \sigma_{A3})\sigma_{B3}\\
    & P(T=7) = \prod_{i=1}^{3}(1 - \sigma_{Ai})(1 - \sigma_{Bi}).
\end{align*}
The pair of players would thus receive the following expected payoffs:
\begin{align*}
    E[\pi_{A/B}] = \sum_{t=1}^{7} \pi_{A/B}(t)\cdot P(T=t).
\end{align*}
We then introduce a second free parameter of our model, $\delta\in[-1,+1]$, to account for other-regarding preferences.
This parameter, usually called \emph{welfare tradeoff ratio}~\citep{QI2022381}, allows to compute the perceived expected payoff, $\Pi$, by taking into consideration how much a player values the co-player’s welfare compared to their own:
\begin{align*}
    \Pi_{A/B} = E[\pi_{A/B}] + \delta E[\pi_{B/A}].
\end{align*}

\subsubsection*{Replicator dynamics}
To understand which strategies are the most successful in our setting, we propose an approach based on Evolutionary Game Theory~\citep{smith_logic_1973,smith_evolution_1982,sigmund_calculus_2010,fernandez_domingos_egttools_2023}.
In this framework, a population of agents repeatedly plays the game with each other, collecting a payoff at each interaction.
Crucially, the agents whose strategy led to a higher reward are imitated by the others.
Over time, less successful strategies are replaced by better ones, until the system potentially reaches a steady state.
In the limit where the population consists of infinitely many players, this process is described by the \emph{replicator equation}~\citep{schuster1983replicator,cressman2014replicator}. 
This deterministic differential equation defines the rate at which the frequency of strategies changes in the population, with strategies with a higher-than-average fitness increasing in frequency while the others decline.

Given that players in the ICG can take two different roles, we adopt here the asymmetric version of the replicator equation~\citep{Accinelli2011}.
If we denote the payoff matrices of player A and player B by the same letters, and by $x_{A}$ and $x_{B}$ the arrays containing the frequency of the strategies of each population, then the system of differential equations takes the following form:
\begin{align*}
    \begin{cases}
        \dot{x}_{A} = x_{A}\left(A\cdot x_{B} -(A\cdot x_{B})\cdot x_{A}^{T} \right)\\
        \dot{x}_{B} = x_{B}\left(x_{A}^{T}\cdot B  - (x_{A}^{T}\cdot B)\cdot x_{B}\right)
    \end{cases}
\end{align*}
Stability (i.e., the property of an equilibrium point to be robust to small perturbations) is achieved in asymmetric replicator dynamics when both $\dot{x}_{A}=0$ and $\dot{x}_{B}=0$. 
Note that every stable steady state is a Nash equilibria, and every Nash equilibrium is a steady state in asymmetric replicator dynamics.

\subsubsection*{Fitting with experimental data}
\begin{figure}[!ht]
\centering
\resizebox{\textwidth}{!}{%
\includegraphics{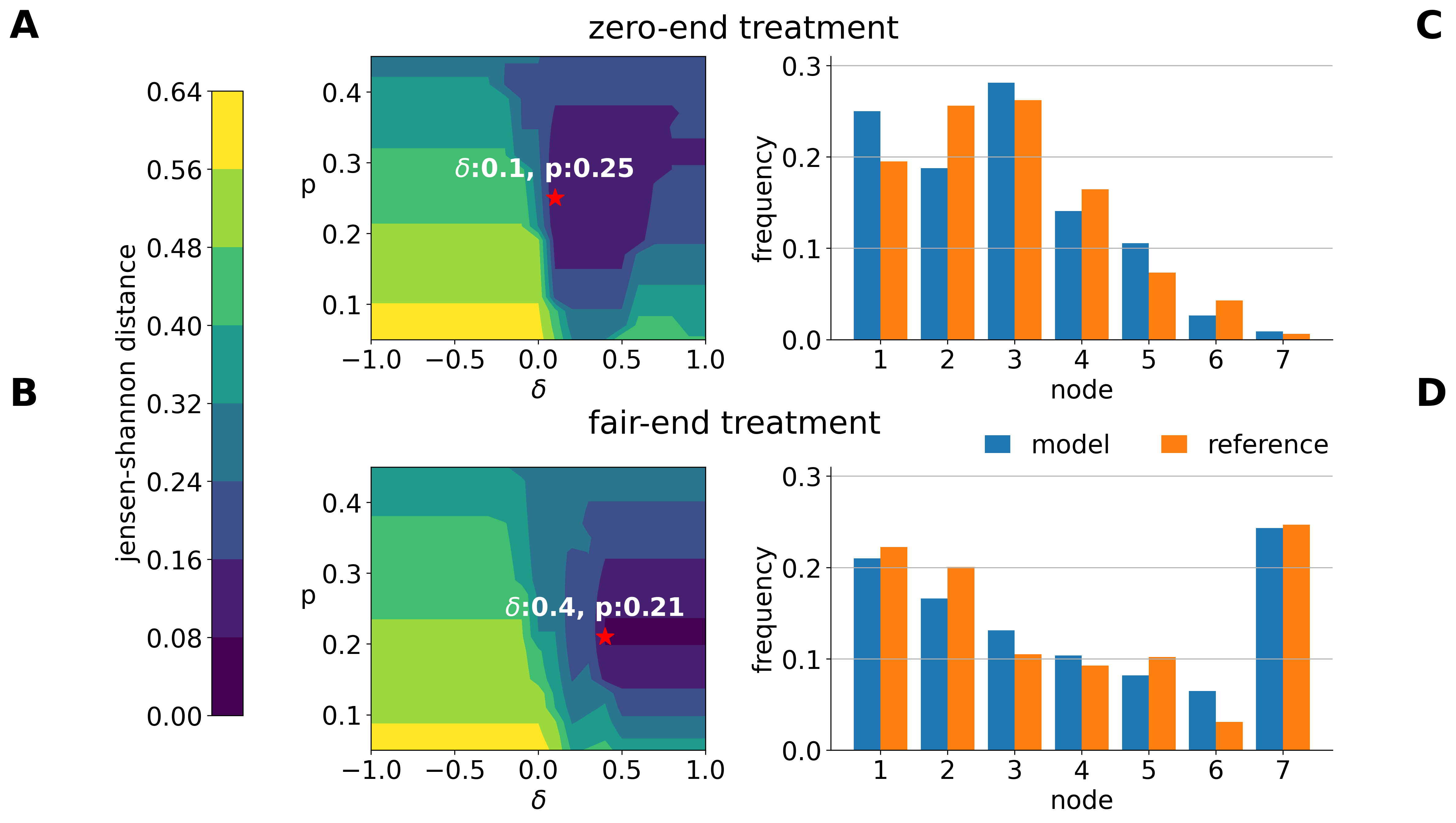}
}
\caption{Fitting of our theoretical model on the experimental data.
Panels \textbf{A, B} show the Jensen-Shannon distance between the prediction of our theoretical model and the experimental reference as a function of the two parameters $p$ and $\delta$, for the zero- and fair-end treatment respectively.
In zero-end treatment, the minimum distance ($0.09$) is achieved when $p=0.25$ and $\delta=0.1$, while in the fair-end treatment minimum distance ($0.07$) is reached when $p=0.21$ and $\delta=0.4$.
Panels \textbf{C,D} show the comparison between the distribution of the terminal node at the optimal fitting and the experimental reference in both cases.
}
\label{fig:fit}
\end{figure}

We address the choice of the pair of values for the parameters of our model, namely the probability value $p$ that specifies the probabilistic strategies and the welfare tradeoff ratio $\delta$ that determines the expected payoffs of our model~\citep{QI2022381}.
For both treatments, the experimental references consist of the distribution of the terminal node averaged over both rounds.
For each pair $(p,\delta)$ in the space $[0,\ 0.5)\times[-1,\ +1]$ we compute the Jensen-Shannon distance between the experimental reference and the terminal node distribution at the Nash equilibrium of the game\footnote{If the game has multiple equilibria, we compute the average distance from the experimental reference.}, separately for each treatment.
The results of the fitting are reported in \autoref{fig:fit}, where the top (bottom) panels refer to the zero- (fair-) end treatment.

The contour plots in Panels \textbf{A,B} show the Jensen-Shannon distance between the outcome distribution predicted by our model and the experimental reference across the parameter space, separately for each treatment. 
The optimal configurations are highlighted in red.
In the zero-end treatment, the minimum distance ($\approx 0.09$) from the experimental data is located at the point $\delta=0.1$, $p=0.25$.
In the fair-end case, the minimum distance ($\approx 0.07$) is obtained at $\delta=0.4$, $p=0.21$.
Our model, tuned with the aforementioned parameters, produces the two terminal nodes' distributions visualised in Panels \textbf{C,D} next to the experimental reference.

\subsection{Statistical models}\label{subsec:statistical-models}

Before fitting predictive models, we assess distributional differences between cohorts to provide a baseline statistical characterisation of the data.
As our data are mostly non-Gaussian, we use the Wilcoxon rank-sum test, also known as Mann–Whitney U test, to perform two-sample comparisons without assuming normality.
Indeed, the Wilcoxon rank-sum test is a non-parametric test for the null hypothesis that randomly selected values $X=\left\{x_{1},\ldots,x_{n}\right\}$ and $Y=\left\{y_{1},\ldots,y_{m}\right\}$ from two independent populations have the same distribution.
Formally, the Wilcoxon rank-sum test relies on the $U-$statistic, defined as $U=\min\left\{nm+\frac{n(n+1)}{2}-R_{X}, nm+\frac{m(m+1)}{2}-R_{Y}\right\}$
$R_{X}$ and $R_{Y}$ are the sums of the ranks in the samples $X$ and $Y$, after ranking all data points from both samples so that the lowest value obtains rank 1 and the highest rank $n+m$.

Then, given the discrete nature of the ICG, we use classification models, which aim at categorizing data into predefined classes (the target) based on their features (the predictors).
Here, we use logistic regression for binary classification (i.e., the target variable can only take two values, usually 0 and 1).
In this model, the log-odds of the event are modelled as a linear function of the predictors $x$, i.e., $\log\left(p/(1-p)\right) = \beta_{0} +\sum_{j}\beta_{j}x_{j}$.

When the outcome consists of ordered categories rather than a simple binary, we use Cumulative Link Models (CLMs). 
These family of statistical models approximate, for each level $j \in \left\{1,\ldots J\right\}$ of the ordinal response $Y$, the cumulative probability of being in level $j$ or lower, $P(Y\le j)$.
CLMs take the general form $g^{-1}\left(P(Y\le j)\right) = \alpha_{j} + X \beta$,
where $g^{-1}$ is the so-called link function, $\alpha_{j}$ is the threshold for each level $j$, $X$ represents the model matrix, $\beta$ the vector of coefficients of the regressors.

The main assumption underlying CLMs is the \emph{proportional odds assumption}, i.e., the effects of any explanatory variables should be the same across different thresholds $\alpha_{j}$.
The assumption means that the coefficients for each predictor category must be consistent (having parallel slopes) across all levels of the response.
If this assumption is violated by certain predictors, a Partial Proportional Odds Model (PPOM) is a valid alternative, as it allows such predictor variables to have a different effect for each level of the output variables.

The goodness of fit of two competing statistical models is compared using the likelihood-ratio test.
This test can be used to evaluate whether the inclusion of additional parameters significantly improves the model fit.
The two models are required to be nested, i.e., the more complex model can be transformed into the simpler model by imposing constraints on the former's parameters.
If the more constrained model (the null model $H_{0}$) is a reasonable explanation for the data, the difference with the better fit of the more flexible model (the alternative model $H_{1}$) should not be significantly higher than the level expected from random sampling variability.
Formally, the likelihood ratio test uses a statistic based on the difference between the log-likelihoods of the two models, $-2\log\Lambda = -2(\log L_{0} - \log L_{1})$.

\subsection{Limitations of the study}
Previous studies have suggested that, although substantial improvements have been made in the market of online platforms for behavioural research, data quality remains a persistent concern for researchers~\citep{peer2022data}.
The choice of the crowdsourcing platform and the implementation of rigorous quality controls are therefore crucial aspects within the design of a behavioural experiment.
Prolific has been widely validated by evaluation surveys, being one of the most reliable crowdsourcing platforms currently available~\citep{douglas2023data,albert2023comparing}.
However, more recent publications have raised general concerns that the validity of online survey research may be undermined by the increasing use of autonomous systems based on technologies such as large language models~\citep{westwood2025potential}.
While all crowdsourcing platforms need to urgently implement policies to guarantee the reliability of the experimental data, researchers can also take preventive measures.
In our case, the comprehension test to assess the understanding of the ICG served as a major barrier for autonomous bots.
In addition, we disabled text selection as well as copy and paste on all pages of our experiment.
Overall, we believe that these countermeasures allowed us to gather reliable human data.
However, we are aware that detecting and avoiding AI-assisted decision-making (e.g., participants who use a second device to ask a chatbot for support on the fly) is much more complicated.

\subsection{Data accessibility}
The data and code implementations on which the manuscript is based on are available at the following (\href{https://anonymous.4open.science/r/fair-cg-4525/README.md}{anonymous link}).
The behavioural experiment was designed using oTree (v. 5.10.4)~\citep{chen2016otree}, an open-source platform for web-based interactive tasks, and then deployed to Heroku, a server provider.
The statistical models were developed on R, using the \emph{ordinal} library~\citep{ordinalref}. The related tables were created with the \emph{stargazer} library~\citep{Hlavac2022stargazer}.
The game-theoretical model was implemented on Python mostly using the \emph{nashpy} library~\citep{knight2018nashpy}.




\clearpage 





\appendix

\section{Tables}\label{appendix:tables}

\begin{table}[pos=htbp]
\centering
\caption{Estimated coefficients of the Partial Proportional Odds (PPO) model.}
\begin{tabular}{l|l|l|l|l}
                   & Estimate & Std. Error & z value  & Pr$(>|z|)$ \\ \hline
1|2.(Intercept)    & \textbf{-1.55501} & 0.38037    & -4.08818 & 0.00004                 \\
2|3.(Intercept)    & \textbf{-0.29946} & 0.35754    & -0.83758 & 0.40227               \\
3|4.(Intercept)    & \textbf{0.83847}  & 0.36175    & 2.31785  & 0.02046               \\
4|5.(Intercept)    & \textbf{1.9141}   & 0.39462    & 4.85049  & 0.00000                     \\
5|6.(Intercept)    & \textbf{2.90936}  & 0.47866    & 6.07808  & 0.00000                     \\
6|7.(Intercept)    & \textbf{5.03134}  & 1.05093    & 4.78752  & 0.00000                     \\ \hline
1|2.treatment (1=F.E.) & \textbf{0.11978}  & 0.27988    & 0.42797  & 0.66868               \\
2|3.treatment (1=F.E.) & \textbf{-0.19758} & 0.23078    & -0.85612 & 0.39193               \\
3|4.treatment (1=F.E.) & \textbf{-0.88344} & 0.24116    & -3.66335 & 0.00025               \\
4|5.treatment (1=F.E.) & \textbf{-1.55232} & 0.29453    & -5.27057 & 0.00000                     \\
5|6.treatment (1=F.E.) & \textbf{-2.06616} & 0.40727    & -5.07317 & 0.00000                     \\
6|7.treatment (1=F.E.) & \textbf{-4.01745} & 1.02145    & -3.9331  & 0.00000                 \\ \hline
round (1=R2) & \textbf{0.31306}  & 0.22906    & 1.36674  & 0.17171               \\
strangers (1=strangers) & \textbf{-0.05995} & 0.28752    & -0.20852 & 0.83482               \\
age$_A$ (scaled) & \textbf{-0.00854} & 0.10046    & -0.08504 & 0.93223               \\
age$_B$ (scaled) & \textbf{-0.19135} & 0.10065    & -1.90117 & 0.05728               \\
sex$_A$ (1=Female) & \textbf{-0.43404} & 0.20135    & -2.1556  & 0.03111               \\
sex$_B$ (1=Female) & \textbf{-0.05874} & 0.20086    & -0.29246 & 0.76993              
\end{tabular}
\label{tab:ppom}
\end{table}

\begin{table}[pos=htbp]
\centering
\caption{Average marginal effects of the treatment on the terminal node of the Incremental Centipede Game (with 95\% bootstrap CI).}
\resizebox{1.\textwidth}{!}{%
\begin{tabular}{cccccccc}
 & 1 & 2 & 3 & 4 & 5 & 6 & 7 \\ 
  Z.E. & 0.198 [0.139, 0.260] & 0.259 [0.195, 0.327] & 0.261 [0.195, 0.331] & 0.162 [0.107, 0.218] & 0.071 [0.036, 0.114] & 0.042 [0.013, 0.073] & 0.006 [0.006, 0.024] \\ 
  F.E. & 0.218 [0.156, 0.283] & 0.193 [0.132, 0.258] & 0.109 [0.063, 0.160] & 0.097 [0.052, 0.144] & 0.104 [0.059, 0.153] & 0.032 [0.007, 0.064] & 0.248 [0.182, 0.320] \\ 
  $\Delta$ & 0.019 [-0.066, 0.107] & -0.067 [-0.158, 0.027] & -0.152 [-0.240, -0.068] & -0.065 [-0.136, 0.009] & 0.032 [-0.030, 0.094] & -0.010 [-0.051, 0.032] & 0.242 [0.171, 0.311] \\ 
\end{tabular}
}
\label{tab:ame}
\end{table}

\begin{table}[pos=htbp]
    \centering 
  \caption{Summary of the logistic model associating behavioural changes between rounds to the participants' features, separated by treatment. For each predictor, we report the estimated coefficient, its standard deviation in round brackets, as well as the Wald test statistic and the p-value.}  
\begin{tabular}{@{\extracolsep{5pt}}lcc} 
\\[-1.8ex]\hline 
\hline \\[-1.8ex] 
 & \multicolumn{2}{c}{\textit{Dependent variable:}} \\ 
\cline{2-3} 
\\[-1.8ex] & \multicolumn{2}{c}{$y=\begin{cases}0\text{ if \# Pass in R1= \# Pass in R2},\\ 1\text{ otherwise}\end{cases}$} \\ 
 & zero-end treatment & fair-end treatment \\ 
\\[-1.8ex] & (1) & (2)\\ 
\hline \\[-1.8ex] 
  role (1=B) & 0.080 & 0.401 \\ 
  & (0.520) & (0.606) \\ 
  & t = 0.154 & t = 0.661 \\ 
  & p = 0.878 & p = 0.509 \\ 
  ended R1 & $-$2.176$^{***}$ & $-$3.014$^{***}$ \\ 
  & (0.539) & (0.814) \\ 
  & t = $-$4.033 & t = $-$3.704 \\ 
  & p = 0.0001 & p = 0.0003 \\ 
  strangers (1=strangers) & $-$0.009 & 0.029 \\ 
  & (0.542) & (0.630) \\ 
  & t = $-$0.017 & t = 0.046 \\ 
  & p = 0.987 & p = 0.964 \\ 
  Sex (1=Female) & 0.193 & $-$0.633 \\ 
  & (0.544) & (0.563) \\ 
  & t = 0.355 & t = $-$1.123 \\ 
  & p = 0.723 & p = 0.262 \\ 
  Age (scaled) & $-$0.282 & 0.496 \\ 
  & (0.250) & (0.322) \\ 
  & t = $-$1.130 & t = 1.540 \\ 
  & p = 0.259 & p = 0.124 \\ 
  Intercept & 1.096$^{*}$ & 2.727$^{***}$ \\ 
  & (0.576) & (0.946) \\ 
  & t = 1.904 & t = 2.884 \\ 
  & p = 0.057 & p = 0.004 \\ 
 \hline \\[-1.8ex] 
Observations & 82 & 78 \\ 
Log Likelihood & $-$45.718 & $-$38.300 \\ 
Akaike Inf. Crit. & 103.435 & 88.600 \\ 
Residual Deviance & 91.435 (df = 76) & 76.600 (df = 72) \\ 
Null Deviance & 113.237 (df = 81) & 102.945 (df = 77) \\ 
\hline 
\hline \\[-1.8ex] 
\textit{Note:}  & \multicolumn{2}{r}{$^{*}$p$<$0.1; $^{**}$p$<$0.05; $^{***}$p$<$0.01} \\ 
\end{tabular} 
\label{tab:model2}
\end{table}

\begin{table}[pos=htbp] \centering 
  \caption{Extension of the PPOM presented in \autoref{tab:ppom} to include two additional continuous predictors ($SVO_A$ and $SVO_B$). For each predictor, we report the estimated coefficient, its standard deviation in round brackets, as well as the Wald test statistic and the p-value.}
\begin{tabular}{@{\extracolsep{5pt}}lc} 
\\[-1.8ex]\hline 
\hline \\[-1.8ex] 
 & \multicolumn{1}{c}{\textit{Dependent variable:}} \\ 
\cline{2-2} 
\\[-1.8ex] & terminal node $\in\left\{1,2,\ldots,7\right\}$ \\ 
\hline \\[-1.8ex] 
 round (1=R2) & 0.379 \\ 
  & (0.231) \\ 
  & t = 1.641 \\ 
  & p = 0.101 \\ 
  strangers (1=strangers) & 0.037 \\ 
  & (0.289) \\ 
  & t = 0.127 \\ 
  & p = 0.899 \\ 
  age$_A$ (scaled) & $-$0.079 \\ 
  & (0.104) \\ 
  & t = $-$0.761 \\ 
  & p = 0.447 \\ 
  age$_B$ (scaled) & $-$0.233$^{**}$ \\ 
  & (0.102) \\ 
  & t = $-$2.295 \\ 
  & p = 0.022 \\ 
  sex$_A$ (1=Female) & $-$0.395$^{*}$ \\ 
  & (0.202) \\ 
  & t = $-$1.952 \\ 
  & p = 0.051 \\ 
  sex$_B$ (1=Female) & $-$0.156 \\ 
  & (0.203) \\ 
  & t = $-$0.768 \\ 
  & p = 0.443 \\ 
  svo$_A$ & 0.224$^{**}$ \\ 
  & (0.107) \\ 
  & t = 2.103 \\ 
  & p = 0.036 \\ 
  svo$_B$ & 0.284$^{***}$ \\ 
  & (0.105) \\ 
  & t = 2.694 \\ 
  & p = 0.008 \\ 
 \hline \\[-1.8ex] 
Observations & 319 \\ 
Log Likelihood & $-$542.885 \\ 
\hline 
\hline \\[-1.8ex] 
\textit{Note:}  & \multicolumn{1}{r}{$^{*}$p$<$0.1; $^{**}$p$<$0.05; $^{***}$p$<$0.01} \\ 
\end{tabular} 
\label{tab:model3} 
\end{table} 
\clearpage

\begin{table}[pos=htbp] 
\centering 
\caption{Extension of the PPOM presented in \autoref{tab:ppom} to include two additional binary predictors ($CRT_A$ and $CRT_B$). For each predictor, we report the estimated coefficient, its standard deviation in round brackets, as well as the Wald test statistic and the p-value.} 
\begin{tabular}{@{\extracolsep{5pt}}lc} 
\\[-1.8ex]\hline 
\hline \\[-1.8ex] 
 & \multicolumn{1}{c}{\textit{Dependent variable:}} \\ 
\cline{2-2} 
\\[-1.8ex] & terminal node $\in\left\{1,2,\ldots,7\right\}$ \\ 
\hline \\[-1.8ex] 
 round (1=R2) & 0.304 \\ 
  & (0.230) \\ 
  & t = 1.324 \\ 
  & p = 0.186 \\ 
  strangers (1=strangers) & $-$0.105 \\ 
  & (0.291) \\ 
  & t = $-$0.360 \\ 
  & p = 0.719 \\ 
  age$_A$ (scaled) & 0.002 \\ 
  & (0.101) \\ 
  & t = 0.015 \\ 
  & p = 0.989 \\ 
  age$_B$ (scaled) & $-$0.182$^{*}$ \\ 
  & (0.101) \\ 
  & t = $-$1.805 \\ 
  & p = 0.072 \\ 
  sex$_A$ (1=Female) & $-$0.432$^{**}$ \\ 
  & (0.202) \\ 
  & t = $-$2.140 \\ 
  & p = 0.033 \\ 
  sex$_B$ (1=Female) & $-$0.026 \\ 
  & (0.207) \\ 
  & t = $-$0.128 \\ 
  & p = 0.898 \\ 
  crt$_A$ (1=reflective) & 0.354$^{*}$ \\ 
  & (0.208) \\ 
  & t = 1.706 \\ 
  & p = 0.089 \\ 
  crt$_B$ (1=reflective) & 0.028 \\ 
  & (0.209) \\ 
  & t = 0.136 \\ 
  & p = 0.893 \\ 
 \hline \\[-1.8ex] 
Observations & 319 \\ 
Log Likelihood & $-$547.439 \\ 
\hline 
\hline \\[-1.8ex] 
\textit{Note:}  & \multicolumn{1}{r}{$^{*}$p$<$0.1; $^{**}$p$<$0.05; $^{***}$p$<$0.01} \\ 
\end{tabular} 
\label{tab:crt} 
\end{table} 

\clearpage
\section{Deviations from the preregistration}\label{appendix:deviations}
Here we report some updates from the preregistered analysis on the OSF. 
Regarding the primary hypothesis, we used a partial proportional odds model (PPOM) instead of a cumulative link model (CLM), as the binary variable associated with the treatment violated the proportional odds assumption.
In the secondary hypothesis, we reduced the complexity of the analysis proposed in the preregistration, as the target variable was unbalanced.
We thus aggregated the data to obtain a relatively balanced binary target to fit a logistic model.

\section{Survey}\label{appendix:survey}

\begin{figure}[pos=htbp]
\centering
\resizebox{0.9\textwidth}{!}{%
\includegraphics{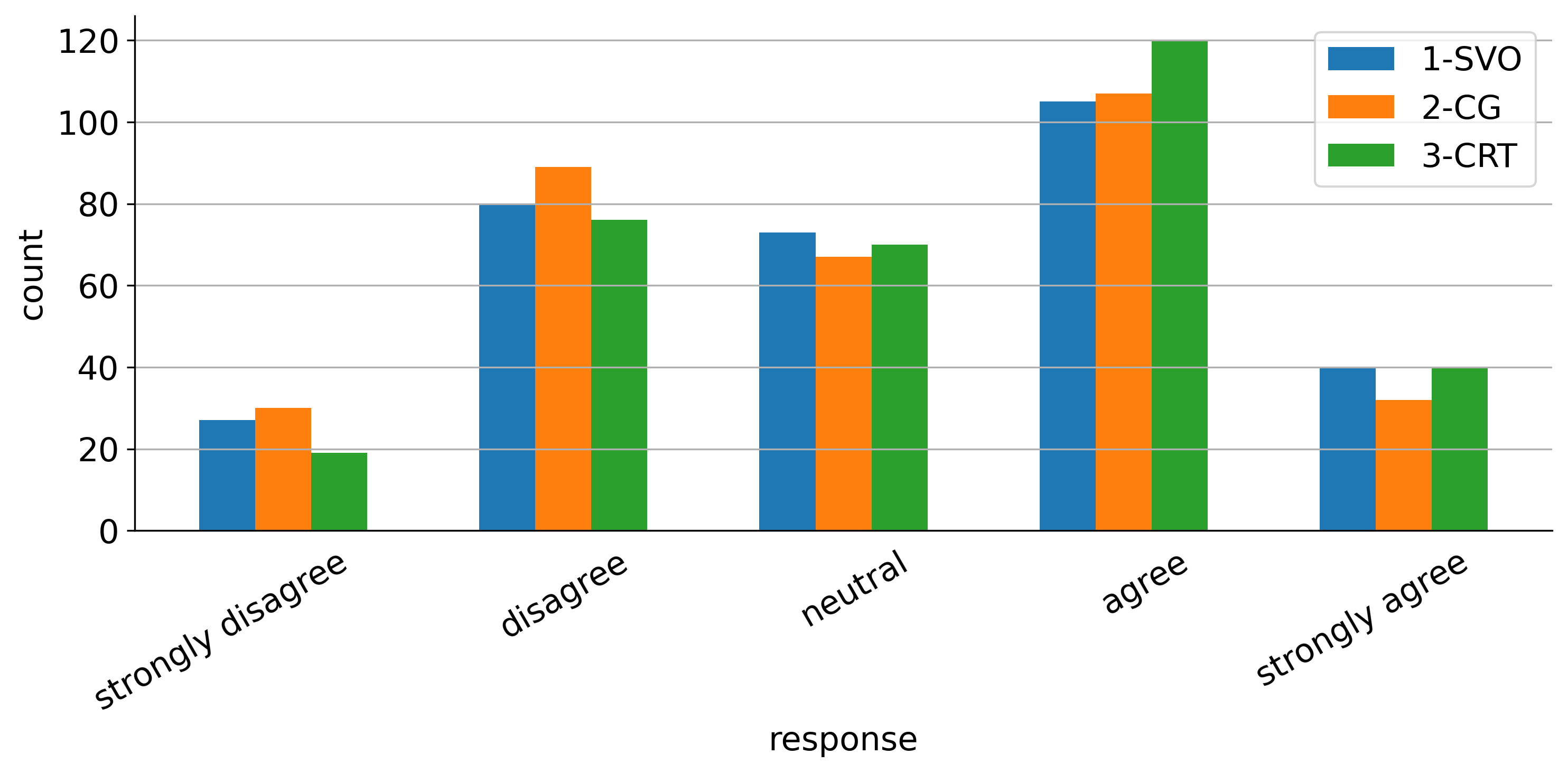}
}
\caption{Familiarity with the tasks of our experiment.
The bar chart reports the counts related to the answers to the questionnaire ``You are familiar with [Task 1/the game you played in Task 2/the questions asked in Task 3]."
Participants could choose their answer on a 5-option Likert scale.
}
\label{fig:survey1}
\end{figure}

\begin{figure}[pos=htbp]
\centering
\resizebox{0.9\textwidth}{!}{%
\includegraphics{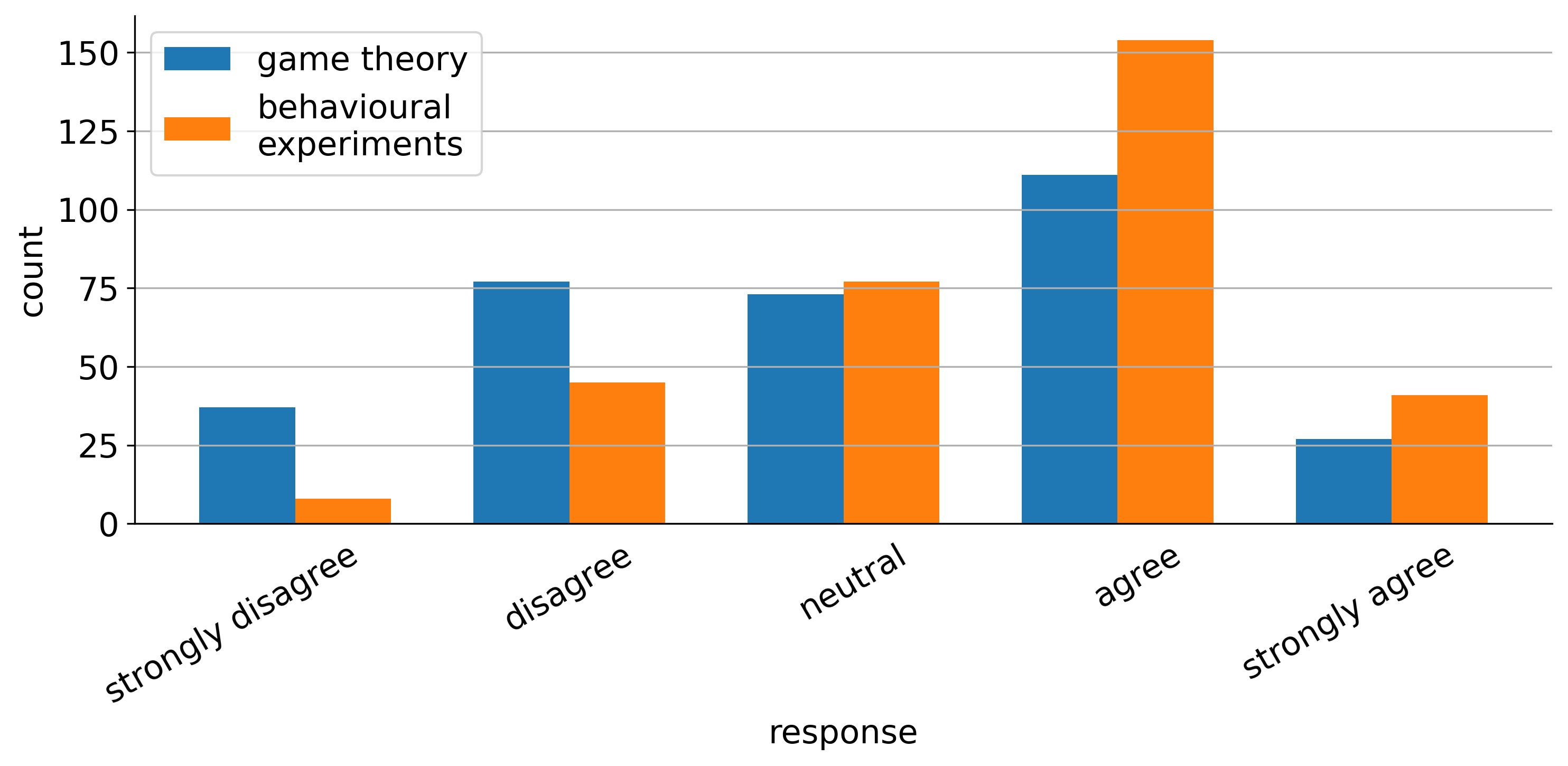}
}
\caption{Familiarity with game theory and behavioural experiment.
The bar plot reports the count of the answers to the questionnaire ``You are familiar with [game theory/behavioural experiments]."
Participants could select their answer on a 5-option Likert scale.
}
\label{fig:survey2}
\end{figure}

\begin{figure}[pos=htbp]
\centering
\resizebox{0.95\textwidth}{!}{%
\includegraphics{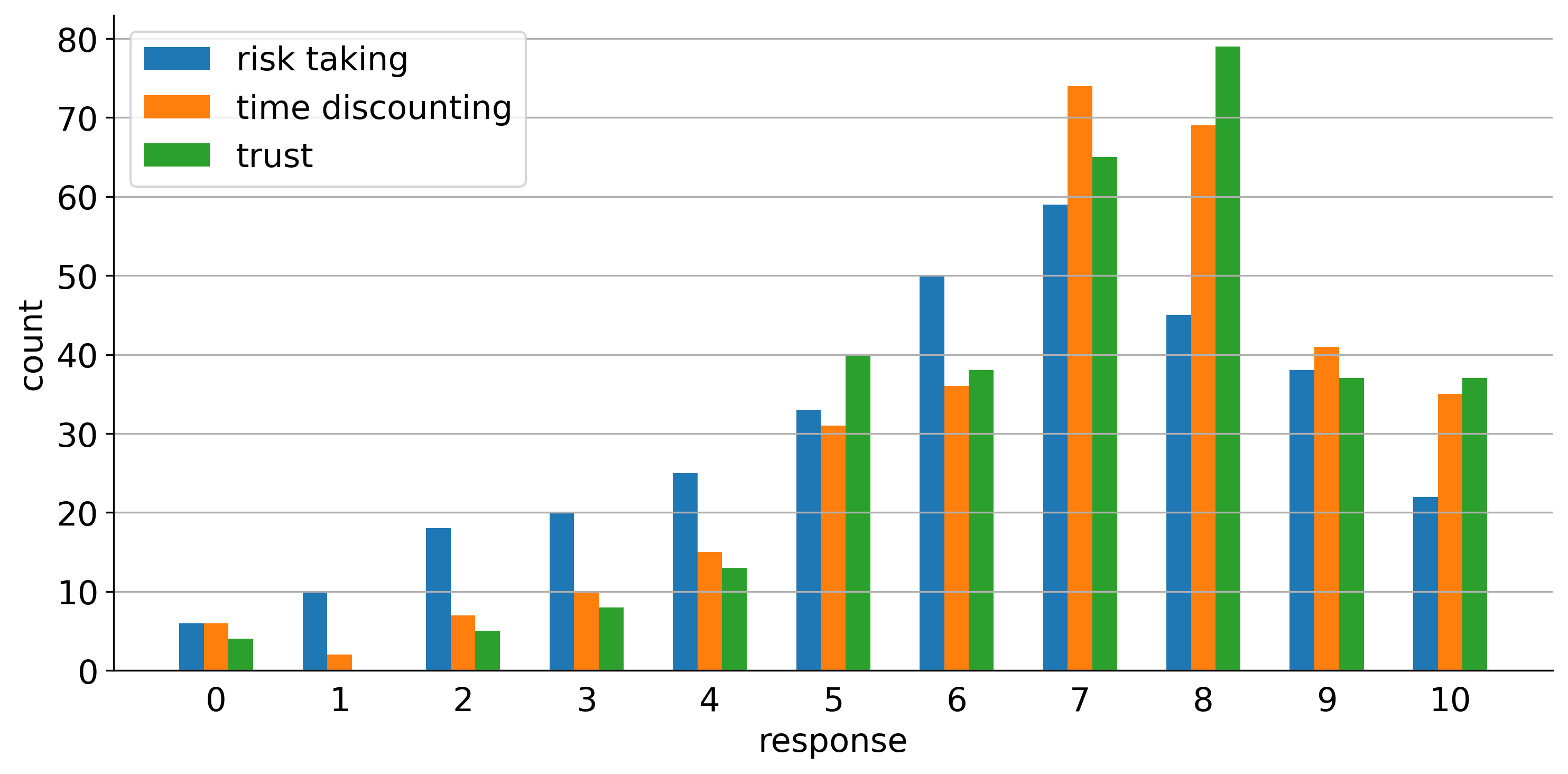}
}
\caption{Qualitative assessment of risk-taking, time discounting and trust.
The risk-taking attitude is assessed through the following question: ``How do you see yourself: are you a person who is generally willing to take risks, or do you try to avoid taking risks?" (0 means ``completely unwilling to take risks", and 10 means ``very willing to take risks").
Time discounting is assessed via the following question: ``In comparison to others, are you a person who is generally willing to give up something today in order to benefit from that in the future or are you not willing to do so?"
(0 means ``completely unwilling to give up something today" and 10 means ``very willing to give up something today").
Finally, trust is measured with the following question: ``How do you assess your willingness to share with others without expecting anything in return when it comes to charity?"
(0 means ``completely unwilling to share" and 10 means ``very willing to share").
}
\label{fig:survey3}
\end{figure}

\clearpage
\section{Demographics}\label{appendix:demographics}

\begin{table}[pos=htbp]
\centering
\resizebox{\textwidth}{!}{%
\begin{tabular}{cc|ccc|ccc|ccc|ccc}
\multicolumn{2}{c|}{Sex}           & \multicolumn{3}{c|}{Ethnicity}                                   & \multicolumn{3}{c|}{Country of birth}                         & \multicolumn{3}{c|}{Country of residence}                     & \multicolumn{3}{c}{Nationality}                                \\ \hline
\multicolumn{1}{c|}{Female} & Male & \multicolumn{1}{c|}{White} & \multicolumn{1}{c|}{Black} & Others & \multicolumn{1}{c|}{UK}   & \multicolumn{1}{c|}{US}  & Others & \multicolumn{1}{c|}{UK}   & \multicolumn{1}{c|}{US}  & Others & \multicolumn{1}{c|}{UK}   & \multicolumn{1}{c|}{US}   & Others \\ \hline
\multicolumn{1}{c|}{0.5}    & 0.5  & \multicolumn{1}{c|}{0.63}  & \multicolumn{1}{c|}{0.22}  & 0.15   & \multicolumn{1}{c|}{0.48} & \multicolumn{1}{c|}{0.2} & 0.32   & \multicolumn{1}{c|}{0.66} & \multicolumn{1}{c|}{0.2} & 0.14   & \multicolumn{1}{c|}{0.54} & \multicolumn{1}{c|}{0.18} & 0.28  
\end{tabular}%
}
\caption{Demographics of our experimental sample. 
The table includes information related to sex, ethnicity, country of birth, country of residence and nationality of the participants.}
\label{tab:demographics1}
\end{table}

\begin{table}[pos=htbp]
\centering
\resizebox{0.7\textwidth}{!}{%
\begin{tabular}{l|r|r|r|r|r|r|r}
 & mean & std & min & 25\% & 50\% & 75\% & max \\ \hline
Time taken (m) & 24.0 & 10.3 & 8.9 & 16.2 & 21.6 & 28.9 & 82.4 \\ \hline
Total approvals & 276.6 & 132.1 & 11.0 & 170.0 & 282.0 & 390.5 & 499.0 \\ \hline
Age & 36.3 & 12.2 & 18.0 & 26.0 & 34.0 & 44.0 & 70.0
\end{tabular}
}
\caption{Information related to our experimental sample.
The table includes information related to time taken to complete the experiment, the total number of approvals and the age of of the participants.}
\label{tab:demographics2}
\end{table}

\section{Experiment material}\label{appendix:experiment-material}

\subsection*{Informed Consent Form}
This form gives you information about the scientific study we are conducting, in order for you to make an informed decision about whether you consent hereto and wish to participate.
We encourage you to consider the information carefully.
If you have questions, please ask them to the researchers via the Prolific platform.
You can also ask questions after the experiment is finished by sending an email to mlgexp@ulb.be.
If you decide to participate, you will be asked to agree to the terms listed in this form and stay to the end of the experiment.
However, if you do not feel comfortable you can leave before the experiment starts by not agreeing to this consent form.

\subsubsection*{What is the purpose of this study?}

This experiment studies how individuals make decisions.
Through a web browser, you will be asked to make decisions within the context of strategic games
involving other participants.
The choices you make in the experiment will directly determine the amount of money you earn.
We do not expect any specific behavior from you.
\textbf{Deceiving participants is strictly forbidden in this experiment.}
Thus, participants will have all the information they require to complete the experiment.

\subsubsection*{Who organized this study?}

This experiment is jointly conducted by the \textbf{MASKED} and \textbf{MASKED}.
\textbf{These universities are the controllers of the data and will be receiving the data to use and retain it for the purposes of scientific research.}
This research project has been examined and accepted by the ethical committee of \textbf{MASKED} on 20/02/2024.
Moreover, all the experiments we perform are done in accordance with the General Data Protection
Regulation (GDPR) and follow the practice and protocols traditionally followed in experimental economics.
As a `data subject’ the GDPR grants you several rights that you can exercise over your personal data:
i) you have the right to access and correct your data, you also have the right to erase your data, to limit their processing, to object to their processing and to transfer your data to third parties; ii) you have the right to withdraw your consent to the processing of your data at any time.
The withdrawal of consent does not affect the lawfulness of the processing of the data obtained prior to the withdrawal of consent.
For more information, you can contact the data protection officers of the ULB (rgpd@ulb.be).
If you feel your personal data has been handled incorrectly, you can always contact the supervisory authority, the Belgian Data Protection Authority
(https://www.dataprotectionauthority.be/) at contact@apd-gba.be.

\subsubsection*{How long will it last and how much will you earn?}

The experiment will last approximately \textbf{30 minutes}.
Your remuneration is composed of a fixed part (a show-up fee of \textbf{\textsterling 3.00 paid upon completion of the experiment}) and a variable part which will be explained later.
Prolific handles the variable part as a bonus payment.

\subsubsection*{What kind of data will be recorded?}

The Prolific platform collects some demographic data about you, which we use for a pre-screening.
These data only consist of gender, age, and geographic location, and we use it to make sure that our experimental sample is balanced and representative.
Moreover, \textbf{during this experiment we will collect data about the decisions you make and how you make them}.
Finally, at the end of the experiment we ask you to fill a short survey about your experience.
No other data will be collected.
It is important to note that once you get paid, we will drop the information about your Prolific ID from the dataset.
Then, \textbf{the data collected throughout this experiment cannot be linked back to you}.

\subsubsection*{Privacy and confidentiality}

As researchers, \textbf{we have a duty of confidentiality regarding the data we collect.}
This means that we undertake, for example in the context of a publication or a conference, never to reveal your name or any other data that could identify you.
Moreover, we may only use the data collected for scientific purposes.
The data collected is pseudo-anonymous in the first phase because Prolific associates the data collected with your Prolific ID.
However, we only use this Prolific ID to handle your payment.
As soon as you get paid, we will erase your Prolific ID from the final dataset.
Thus, in this final dataset it is no longer possible to identify you directly.
Once the data has been processed completely anonymously, the GDPR no longer applies.

\subsubsection*{Data retention time}

The fact that you participated in the experiment will be kept confidential and will be known only to the Prolific platform.
You have a right to withdraw consent to that information at any time, and you may do so through the Prolific platform, which will handle this request.
Once the data from this experiment is downloaded from Prolific, it will be fully anonymized,
and all Prolific IDs are eliminated.
This anonymized dataset produced from this experiment will be retained by the \textbf{MASKED} and by \textbf{MASKED} for research purposes, and it will be made public to the scientific community.
Once more, \textbf{the data present in this dataset cannot be linked back to you.}
The remaining data of the experiment in Prolific will be retained until the experiment is published.
During this retention time, you may exercise your right to erase this data by making a direct request to Prolific.

\subsubsection*{Is there a health risk or risk of discomfort?}

No, \textbf{there are no health risks}, and \textbf{the risk of discomfort is very small}.
Most participants enjoy the experience.
However, you might get disappointed with the amount of money you earn and/or other people’s behavior.
Please contact us with any questions or concerns following the experiment via mlgexp@ulb.be.

\subsubsection*{Participant consent}
By signing this document you give your consent, and you indicate that you understand your rights, agree with the above conditions, and agree to participate in the experiment.
Specifically, you agree to the following statements:
\begin{enumerate}
    \item I declare that I am informed about the nature, purpose, duration, potential benefits, and risks of the study and that I know what is expected of me.
    \item I have had enough time to consider the conditions of the study carefully, I have been able to ask all the questions I have, and I have received a clear answer to my questions.
    \item I understand that my participation in this study is voluntary and that I am free to stop my participation in this study without providing a reason.
    \item I understand that during my participation pseudo-anonymized data about me, handled by Prolific, will be collected and that the researcher ensures the confidentiality of this data in accordance with the relevant Belgian and European privacy legislation (Cf. AVG or GDPR).
    \item I agree to the processing of my personal data in accordance with the modalities described in the ``Privacy and confidentiality" section.
    \item I consent to the processing of my data for scientific purposes.
    \item I consent to the publication of the research results.
    My name will not be published, and the confidentiality of the data is guaranteed at every stage of the research.          
\end{enumerate}

$<\text{checkbox}>$ I have read and agree with all the conditions in this consent form.

$<\text{Next}>$

\subsection*{Attention Pledge}

Please give your \textbf{full attention} throughout the experiment.
The experiment typically \textbf{lasts around 30 minutes}.
During the experiment, a timer will be displayed for each task.
We ask you to complete each task within the given time.
Once it begins, please \textbf{remain in the session until the platform confirms that it is complete}.
Do not close or refresh the browser tab, change tabs, or delete the cookies from your
browser during the experiment, as these are necessary to track your progress during the experiment.
If we detect that you have left the browser tab, \textbf{we will issue a warning}.
If you are out for more than 2 minutes, we will consider you as inactive, and you will be removed from the
experiment.
\textbf{Attention: if you do not comply with these rules, you will not be paid.}
Thank you for your cooperation!

$<\text{checkbox}>$ I understand and agree to the conditions stated above.

$<\text{Next}>$

\subsection*{Instructions of the experiment}

You will participate in an experiment on decision-making conducted by the \textbf{MASKED} and \textbf{MASKED}.
\textbf{Everyone receives the same instructions, which contain all the necessary information to complete the experiment, so please read carefully.}
If you have any questions regarding the experiment, use the Prolific chat to contact the researchers.
\textbf{Your privacy is guaranteed.}
Your identity will be kept confidential and results will be anonymous.
\textbf{Your earnings will depend on your choices and the choices of other participants.}
All your earnings during the experiment will be expressed in Experimental Currency Units (ECUs).
Each ECUs is converted to POUNDS (\textsterling ) with an exchange rate of \textbf{100 ECUs = \textsterling 1}
(or \textbf{1 ECU = \textsterling 0.01}).
\textbf{The experiment will last approximately 30 minutes}, and it will include \textbf{3 tasks and a final survey.}

You will receive further instructions before each task.

Your final earnings are calculated by summing a \textbf{fixed amount}
(\textsterling 6/hour, so around \textsterling 3) plus a \textbf{variable amount} (paid as a bonus).
The bonus is calculated as the sum of earnings of all 3 tasks, and it will be a maximum of
\textsterling 2.96.
We will only show you your final earnings at the end of the experiment.

Notes:
\begin{enumerate}
    \item Completing the experiment is necessary to receive the bonus.
    \item Do not close the window, change tabs, refresh the page, or delete your cookies during the experiment to avoid being disqualified.
    \item If you exceed the maximum time assigned to a task, you will be considered to have abandoned the experiment and will not be paid.
\end{enumerate}
 Once you have read these instructions carefully, click ``Next" to proceed to Task 1.
\textbf{Once you click ``Next" you cannot go back.}

$<\text{Next}>$

\subsection*{Task 1 of 3}

\subsubsection*{Instructions}

In this task, you will be presented with \textbf{6 choice situations}.
In each situation, \textbf{you will be asked to split a sum of money between you and another person}.
In each situation, you have to choose between \textbf{9 different alternatives}.
Each alternative involves a distribution of a sum between you and the other person.   
You will have a total of \textbf{15 minutes} to complete this task.
                
\textbf{Example}

The following image illustrates a typical choice situation as it will appear on your screen during the task.

\begin{figure}[!ht]
\centering
\resizebox{\textwidth}{!}{%
\includegraphics{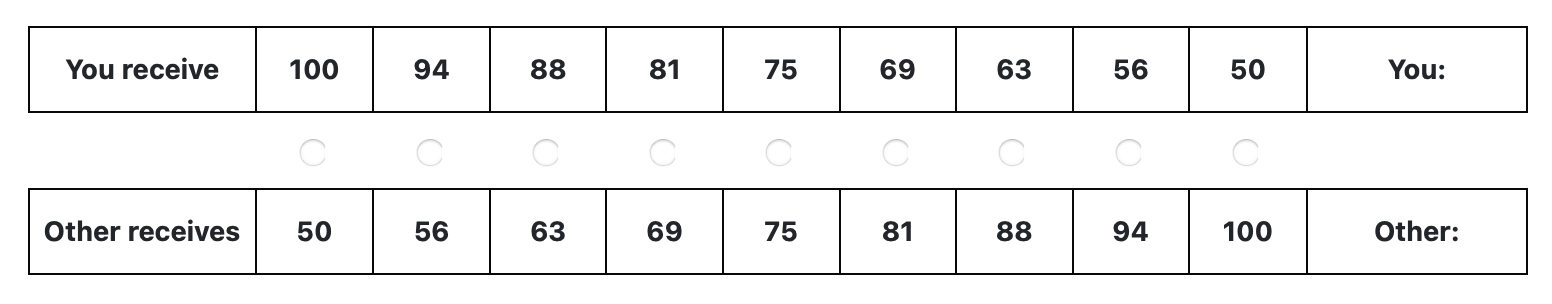}
}
\end{figure}

\begin{enumerate}
    \item In this example, choosing the 3rd option starting from the left would give you 88 ECUs, while the other person would receive 63 ECUs.
    \item On the other hand, choosing for example the 2nd option starting from the right would give you 56 ECUs, while the other participant would receive 94 ECUs.             
\end{enumerate}

\textbf{Making a choice}

You make your choice by clicking on your preferred alternative.
You can change your choice as many times as you wish.
However, once you click the ``Confirm" button at the bottom right, your choice will be final and cannot be changed.
Immediately after, the next choice situation will appear on your screen until you have made all 6 choices.

\textbf{Earnings}

At the end of the study, there is a 1-in-10 chance (10\% chance) to be selected to receive a financial bonus based on this task.
If you are selected, you will be assigned one of two roles: ``Giver" or ``Receiver".

\begin{enumerate}
    \item \textbf{If you are the Giver}: One of the 6 choices you made during the study will be randomly selected.
    You will receive the monetary amount you allocated to yourself in that choice.
    \item \textbf{If you are the Receiver}: Another participant from the study will be randomly selected.
    One of their choices will also be randomly selected, and you will receive the amount they decided to allocate to the ``other" person (you).                           
\end{enumerate}

In both cases, the selection process is entirely random, ensuring fairness and unpredictability in determining payouts.
Make your decisions carefully, as they could directly impact the rewards for yourself or another participant.
The amount of points you earned in Task 1 will be shown on the screen at the end of the experiment.
The points will be converted to POUNDS (\textsterling ) with an exchange rate of \textbf{100 ECUs = \textsterling 1} (or \textbf{1 ECU = \textsterling0.01}).

Once you are ready, please click ``Next" to start Task 1.

$<\text{Next}>$

\subsection*{Task 2 of 3}
\subsubsection*{Instructions 1/3 - Description of the task}

In this task, \textbf{you will play a game for 2 times, each time with a different participant}.

Before the task begins, you will be randomly assigned a role: either \textbf{Player A or Player B}.
Each game pairs a Player A with a Player B.
If you're assigned to role B, for example, you will always be paired with someone in role A (and vice versa).

In each game, \textbf{you and your partner will take turns to decide how to split a sum of money
between the two of you}.

\textbf{Structure of each round}

\begin{enumerate}
    \item Each game lasts a maximum of 6 turns and starts with Player A making the first decision.
    \item At each turn, the player who makes the decision can either choose \textbf{\emph{Take}} or \textbf{\emph{Pass}}.
    \begin{enumerate}
    \item If you choose \textbf{\emph{Take}} the round ends immediately,
        and you and the other participant will receive a payoff according to the split in that turn.
    \item If you choose \textbf{\emph{Pass}} the round continues and the turn shifts to the other participant -  unless you are at the last turn where the final split is applied to both players.
    \end{enumerate}
    \item At each turn, the available split changes.
    \item The payoff for both players will be shown on the screen at the end of each round.
\end{enumerate}

\textbf{Earnings}

Your final payoff for this task is determined as follows:

\begin{enumerate}
\item After all 2 games are played, \textbf{we will randomly select one game to count for your final payment}. Think of it like flipping a coin to pick one of your 2 games.
\item The amount you earned in that selected game will be your final payoff for this task,
and it will be shown at the end of the experiment.
\end{enumerate}

The points will be converted to POUNDS (\textsterling) with an exchange rate of \textbf{100 ECUs = \textsterling 1} (or \textbf{1 ECU = \textsterling 0.01}).

Click on ``Next" to proceed onto the next Instruction page
(you will be able to go back to this page if needed).

$<\text{Next}>$

\subsubsection*{Instructions 2/3 - Example}

The following image illustrates how each game unfolds.
\textbf{The splits shown in the image are the same for every game you will play.}
This scheme will always be available on the screen to help you keep track of who is making the current decision, what the current split is, and what the outcome of each decision would be.

\begin{figure}[!ht]
\centering
\resizebox{\textwidth}{!}{%
\includegraphics{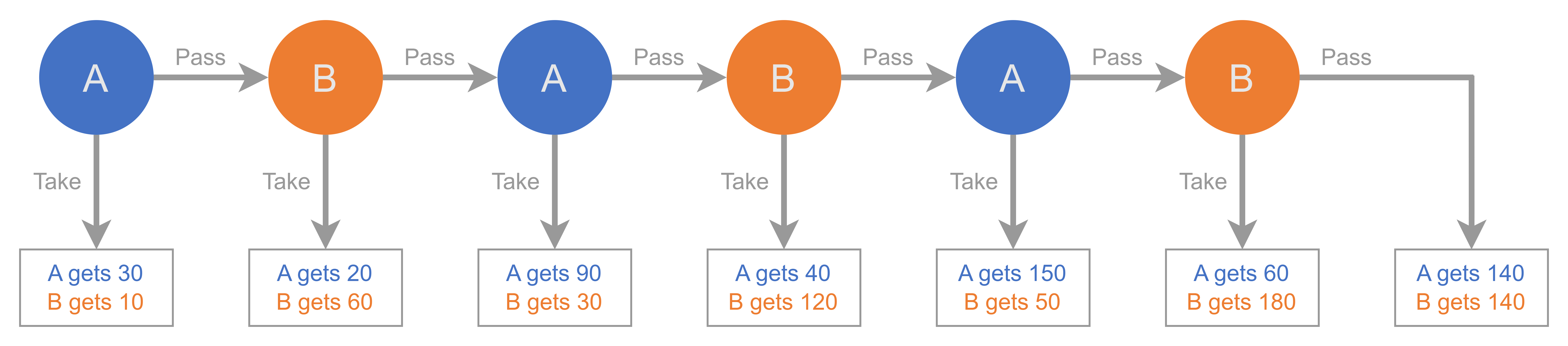}
}
\end{figure}

In the image, each circle represents a turn, and the box below each circle shows the split for that turn.

\begin{enumerate}
\item \textbf{Blue circles} = Player A’s turn
\item \textbf{Orange circles} = Player B’s turn
\end{enumerate}

At each turn, the circle wherein a decision has to be made will be highlighted (see the following example).

\begin{figure}[!ht]
\centering
\resizebox{\textwidth}{!}{%
\includegraphics{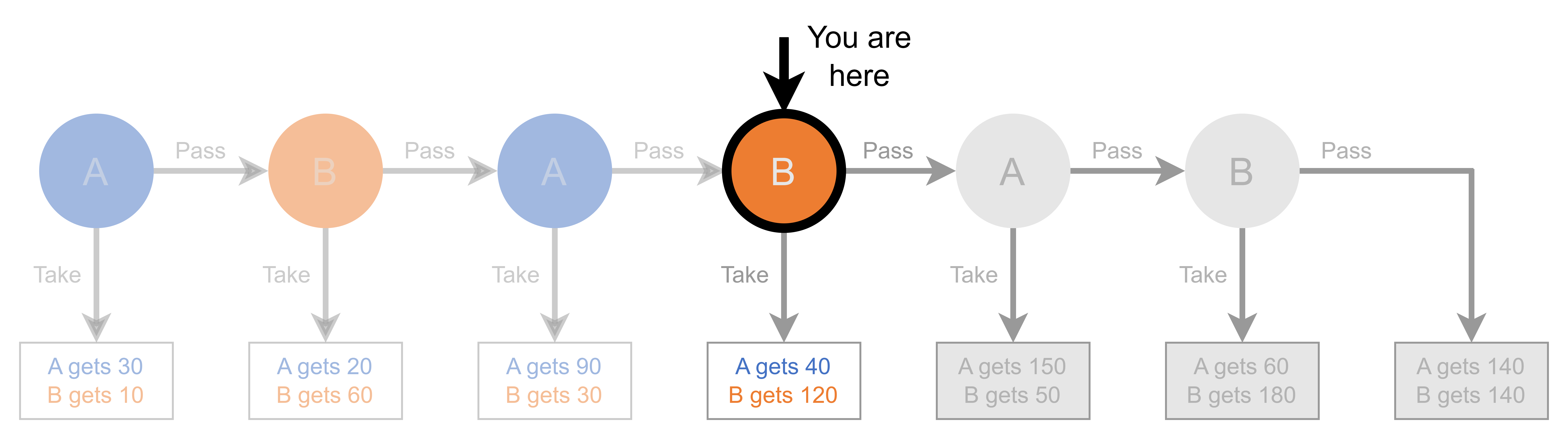}
}
\end{figure}

\begin{enumerate}
\item In this example, on Turn 4, if Player B chooses \textbf{\emph{Take}}, they receive 
120 ECUs and Player A gets 40 ECUs.
\item If Player B chooses \textbf{\emph{Pass}} instead, the game continues to Turn 5, where Player A can choose \textbf{\emph{Take}} (resulting in 150 ECUs for Player A
and 50 ECUs for Player B), or \textbf{\emph{Pass}} again.
\end{enumerate}

Click on ``Next" to proceed onto the next Instruction page
(you will be able to go back to this page if needed).

$<\text{Previous}>$ $<\text{Next}>$

\subsubsection*{Instructions 3/3 - Summary}
\begin{enumerate}
\item Your identity will remain anonymous.
\item All participants receive the same instructions.
\item You will play a total of \textbf{2 games}.
\item You will be randomly assigned a role (\textbf{Player A or Player B})
and you will keep such role during the 2 games.
\item You will be randomly paired with a different participant in each of the 2 games.
\item Player A and Player B alternate turns (up to 6 decisions),
and Player A always makes the first decision.
\item On your turn, you can either choose \textbf{\emph{Take}} to end the game and obtain the current split,
or \textbf{\emph{Pass}} to continue to the next turn.
\item The split changes at each turn.
\item If Player B chooses \textbf{\emph{Pass}} at the 6th turn, the final split is applied to both players.
\item After all games are played, \textbf{we will randomly select one game to count for your final payment}.
\end{enumerate}

\textbf{Important notes}

\begin{enumerate}
\item 
    \textbf{Comprehension Test}:
    You must pass a short comprehension test to proceed.
    You have 5 attempts.
    If you fail 5 times, the experiment will be terminated, and you will not be paid.
\item 
    \textbf{Waiting for other participants}:
    You may wait up to 10 minutes to be paired for your first game.
    This ensures that enough participants are available to allow smooth pairings in your subsequent games.
    Leaving the tab or browser during this time will remove you from the experiment, and you will not be paid.
    If no match is found in time, the experiment will terminate and you will be paid for your time and the previous tasks.
\item 
    \textbf{Decision times}:
    Each decision has a 2-minute limit.
    Timing out results in removal from the experiment, without being paid.
    If your partner gets timed-out during a game,
    you will also be excluded from the experiment,
    but you will be paid for your time and the previous tasks.
\item 
    \textbf{Dropouts}:
    If you drop out at any point, you will not be paid.
    If your partner drops out during a game,
    you will also be excluded from the experiment,
    but you will be paid for your time and the previous tasks.
\end{enumerate}

Click on ``Next" to proceed onto the Comprehension test.
Once you click ``Next” you cannot go back.

$<\text{Previous}>$ $<\text{Next}>$

\subsubsection*{Comprehension test}
\begin{figure}[!ht]
\centering
\resizebox{\textwidth}{!}{%
\includegraphics{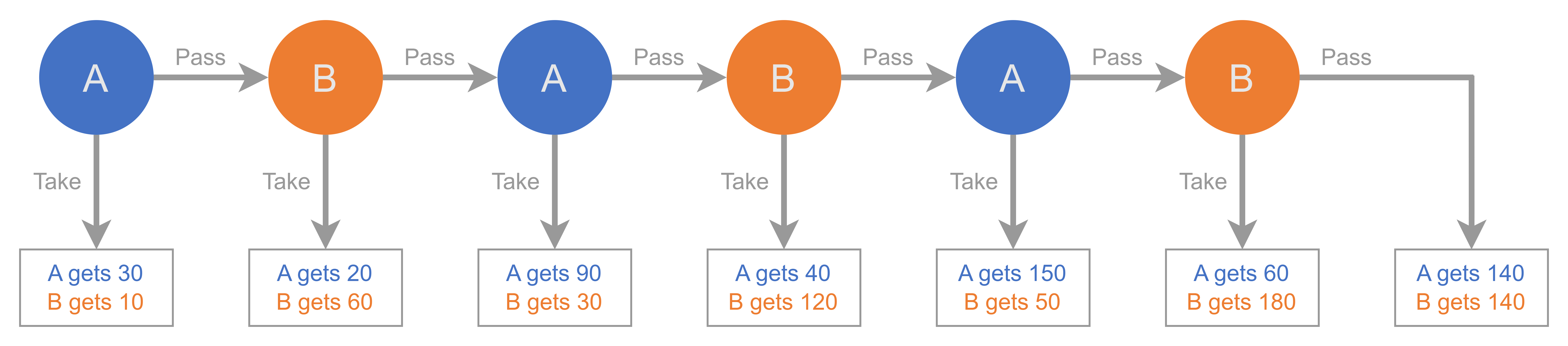}
}
\end{figure}

Please, answer the following questions carefully to test your understanding of the task.
Type numerical characters only.
\textbf{You have 5 attempts to answer them correctly.}
If you do not answer correctly after 5 attempts, you will be removed from the experiment, and you will not receive any of the bonus payments for the previous tasks you completed.

\begin{enumerate}
    \item At the first decision point (\textbf{Step 1}), Player \textbf{A} chooses to \textbf{\emph{Pass}}.
    Then Player \textbf{B} chooses to \textbf{\emph{Take}}.
    How many points does Player \textbf{B} receive?
    \item At the third decision point (\textbf{Step 3}), Player \textbf{A} chooses to \textbf{\emph{Take}}.
    How many points does Player \textbf{A} receive?
    \item At the fourth decision point (\textbf{Step 4}), Player \textbf{B} chooses to \textbf{\emph{Take}}.
    How many points does Player \textbf{A} receive?
    \item At the second decision point (\textbf{Step 2}), Player \textbf{B} chooses to \textbf{\emph{Pass}}.
    Then Player \textbf{A} chooses to \textbf{\emph{Take}}.
    How many points does Player \textbf{B} receive?
    \item At the fifth decision point (\textbf{Step 5}), Player \textbf{A} chooses to \textbf{\emph{Pass}}.
    Then Player \textbf{B} chooses to \textbf{\emph{Pass}} as well.
    How many points does Player \textbf{A} receive?
\end{enumerate}

Once you are ready, please click ``Next" to submit your answers.

$<\text{Next}>$

\subsection*{Task 3 of 3}
\subsubsection*{Instructions}
The purpose of this task is to understand how people reason about simple numerical problems.
You will be presented with 4 consecutive screens, each displaying a question that you have to answer.
\textbf{You will have 1 minute to answer each question.}
After 1 minute, the screen will display the next question.
An example of what a question screen will look like is shown below on \textbf{Figure 2}.

\begin{figure}[!ht]
\centering
\resizebox{\textwidth}{!}{%
\includegraphics{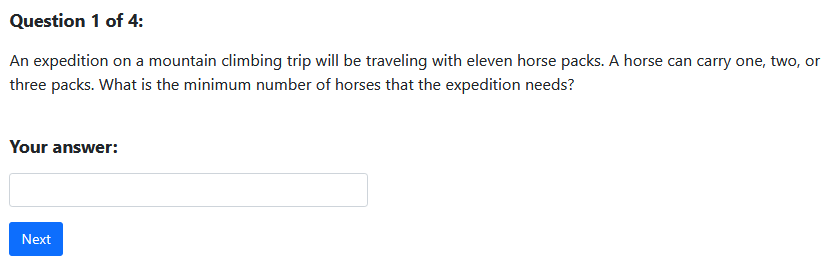}
}
\end{figure}

To answer each question, type your answer into the blank box provided for that question.
\textbf{Please note that you must type numeric characters only.}
Once you have typed your answer, please confirm your choice by clicking "Next".
\textbf{Please note that once you click "Next", your choice is definitive and cannot be changed.}
After each question, there is no feedback about the result.

In this task \textbf{you will receive 5 ECUs for each correct answer.}
All 4 questions are eligible for payment, so if you answer all 4 questions correctly,
you will receive a bonus of 20 ECUs.

The amount of points you earned in Task 3 will be shown on the screen at the end of the experiment.
The points will be converted to POUNDS (\textsterling ) with an exchange rate of \textbf{100 ECUs = \textsterling 1} (or \textbf{1 ECU = \textsterling 0.01}).

Once you are ready, please click "Next" to start Task 3.

$<\text{Next}>$


\printcredits

\bibliographystyle{cas-model2-names}

\bibliography{cas-refs}



\end{document}